\documentclass{article}

\usepackage{arxiv}
\usepackage[utf8]{inputenc} 
\usepackage[T1]{fontenc}    
\usepackage{hyperref}       
\usepackage{url}            
\usepackage{booktabs}       
\usepackage{amsfonts}       
\usepackage{nicefrac}       
\usepackage{microtype}      
\usepackage{lipsum}
\usepackage{epsfig}
\usepackage{epstopdf}
\usepackage{amsmath}
\usepackage{caption}
\usepackage{algorithm}
\usepackage{algorithmicx}
\usepackage{algpseudocode}
\usepackage{ifthen}  
\usepackage{cite} 
\usepackage{url}  
\usepackage{multicol}   
\usepackage{tcolorbox}
\usepackage[utf8]{inputenc}

\title{Block Prefix Mechanism for Flow Mobility in PMIPv6 Based Networks}

\author{
	K Vasu \\
	Department of E \& ECE\\
	IIT Kharagpur\\
	INDIA \\
	\texttt{vasukanster@gmail.com} \\
	\And
	Sudipta Mahapatra \\
	Department of E \& ECE\\
	IIT Kharagpur\\
	INDIA \\
	\texttt{sudipta@ece.iitkgp.ac.in} \\
	\And
	C S Kumar\\
	Department of Mechanical Engineering\\
	IIT Kharagpur\\
	INDIA \\
	\texttt{kumar@mech.iitkgp.ac.in} \\
}

\begin{document}
\maketitle
\vspace{0.5cm}
\begin{abstract}
	The next generation Internet is deemed to be heterogeneous in nature and mobile devices connected to the Internet are expected to be equipped with different wireless network interfaces. As seamless mobility is important in such networks, handover between different network types, called vertical handover, is an important issue in such networks. While proposing standards like Mobile IPv6 (MIPv6) and Proxy Mobile IPv6 (PMIPv6) for mobility management protocols, one important challenge being addressed by IETF work groups and the research community is flow mobility in multi-homed heterogeneous wireless networks. In this paper we propose and analyze a block prefix mechanism for flow mobility in PMIPv6 and conducted extensive analytical and simulation studies to compare the proposed mechanism with existing prefix based mechanisms for flow mobility in PMIPv6 reported in terms of important performance metrics such as handover latency, average hop delay, packet density, signaling cost and packet loss. Both analytical and simulation results demonstrate that the proposed mechanism outperforms the existing flow mobility management procedures using either shared or unique prefixes.
	\end{abstract}
		

\section{Introduction}
As 4G networks are going to be heterogeneous in nature, multi-homing that enables a mobile device to connect to different networks at different times is an essential feature of such networks. One of the important problems in such a scenario is flow mobility, which ensures that the user flow can be moved from one network interface to another network interface as and when it is necessary. In the literature several approaches or methods are identified to address the flow mobility in multi-homed wireless networks \cite{sousaMNA}. 
These approaches are broadly classified into routing based, transmission control based, identifier based and dynamic flow mobility based, distributed and context aware approaches, and finally some enhancements to basic PMIPv6 protocols.\\

A structured Classless Inter-Domain Routing (CIDR) approach for IPv6 multi-homing, which takes care of routing table maintenance, IPv4 address limitation and multi-homing, necessitated by an increase in number of networks and the need for global routing, is proposed by \cite{SenevirathnaICIIS}. Route optimization approaches in flat and distributed architectures of the PMIPv6 domain are explored by (\cite{BocWPMC}, \cite{GoldenbergSIGCOMM}) 
to optimize the network latency, cost, and network. A method using the Mobile access gateway Address Translation (MAT), which can continuously deliver packets even in the case of handover between heterogeneous access technologies, is proposed in \cite{KimICIS}. However, the routing approaches cannot achieve seamless flow mobility in multi-homed wireless networks due to the lack of transport level control. Transmission control approaches, such as Concurrent Multi-path Transfer (CMT) are proposed by \cite{WangCOMCOM} to realize seamless flow mobility. Other transport control approaches include \cite{ChangqiaoIEEETOB}. However, these are not sufficient to solve flow mobility due to the lack of identification of flows from the originating locations, leading to the design of identifier based approaches (\cite{WangCFI}, 
\cite{MappWAINA}, \cite{LijuanEDT}, \cite{ChakchaiCCNC}.

In identifier based approaches various strategies such as ID/Locator split methods, introducing a special identification layer, virtual interfacing in mobile node (MN),and with some special protocols. However, during the flow mobility in heterogeneous wireless networks, various signalling procedures and binding updates are necessary, which ultimately increase the signalling overhead in networks and the load at local mobility anchor (LMA). Moreover, IP data traffic from 3G to WLAN is an essential feature in future networks and energy consumption in a mobile device is one of its critical components. This leads to the development of dynamic flow mobility management approaches (\cite{AlpcanIEEETOMC}, \cite{HongIWCMC}, \cite{MeliaWPC}, \cite{MyungCUTE}, \cite{ShenWPC}, \cite{SunICTC}). In these approaches policies and prefixes are used to dynamically control the flow management in multi-homing scenarios. The dynamic flow mobility management approaches do not consider how to optimistically utilize the resources among interfaces.\\

In the existing resource allocation mechanisms for a heterogeneous wireless access environment supporting mobile terminals (MTs) with multi-homing capabilities a central resource manager provides a global view for resource allocation and admission control. This is not practical if these networks are operated by different service providers. To overcome this problem, a distributed solution for resource allocation and admission control is provided by \cite{IsmailIEEEJOSAC}. A novel distributed solution for network-based localized mobility management in a flat architecture without central mobility anchors is provided by \cite{GiustWPMC}. Although these solutions overcome most of the issues in current centralized architectures, these mechanisms do not ensure end-to-end QoS of various flows in a multi-homed network. Even the recent enhancements to PMIPv6 such as (\cite{JeonIETC}, \cite{KiIEEECL}, \cite{JiWAINA}, \cite{MagagulaEURASIP}, \cite{MurganteLNCS}, \cite{TranINFOCOM}, \cite{YangICC}) also do not provide complete seamless flow mobility in multi-homed heterogeneous wireless networks while optimizing the network latency, signalling cost and QoS of the  end-user application.\\

Moreover, for some of these approaches, the networking infrastructure needs to be changed and one may need to change the protocol stack of the mobile node. So, in this paper we propose a block prefix mechanism to support flow mobility in PMIPv6. The mechanism does not need any change in the existing infrastructure of PMIPv6 domains. Moreover, the analytical and simulation results demonstrate that the proposed method improves flow mobility management in PMIPv6 in terms of handover latency, signalling cost and packet loss. Rest of the paper is organized as follows: Section 2 briefly discusses the related work with example scenarios; Section 3 discusses the proposed block prefix mechanism, followed by theoretical evaluation approaches, both analytical and simulations, in Section 4. Section 5 contains the analytical and simulation results and finally Section 6 draws the conclusions. 

\section{Related work}
The IETF network mobility group has introduced a new mobility management protocol PMIPv6 that eliminates the MN involvement in the handover execution phase. Protocol extensions to PMIPv6 have been introduced to allow flow mobility among the different physical interfaces (\cite{draft-bernardos-netext-pmipv6-flowmob-03}; \cite{draft-sarikaya-netext-flowmob-ext-00}). This can be achieved by properly managing the prefixes, and forwarding policies. In this paper, we address the flow mobility scenarios for single and multi LMA environment.
\subsection{Single LMA}
In the presence of single LMA, flow mobility may be analyzed for two different scenarios:\\ \noindent 1) when all the interfaces are active and when flow mobility takes place among the interfaces,  2) when some of the interfaces are suddenly powered on and the flows need
to be moved from the currently active interfaces to newly active interfaces. Moreover, there are two different prefix management mechanisms proposed to address the flow mobility among these cases \cite{draft-ietf-netext-pmipv6-flowmob-05}: (i) The MN obtains the same prefix or the same set of prefixes as already assigned to an existing session, (ii) The MN obtains a new prefix or a new set of prefixes for the new session (basic PMIPv6). Thus, there are a total of four possible cases: (1) Assigning same set of prefixes to flows when all interfaces are active, (2) Assigning unique set of prefixes to flows when all interfaces are active, (3) Assigning same set of prefixes to flows when some of the interfaces are suddenly powered on, (4) Assigning unique set of prefixes to flows when some of the interfaces are suddenly powered on. In this paper, we use the terms same or common or shared prefixes and the terms unique or different set of prefixes interchangeably.
\subsubsection{Case 1: When all interfaces are active and flows use shared prefix}
When all interfaces are active and flows use shared prefix. In this case no further signalling is required between LMA and MAG because routing entries are already present in both LMA and MAG. All physical interfaces are assigned with same prefix (i.e. pref 1) upon attachment to the MAGs. If the IP layer at the mobile node shows a single interface, then the mobile node has one single IPv6 address , pref1::MN,  configured at the IP layer. Otherwise, three different IPv6 addresses (pref1::IF1, pref1::IF2, pref1::IF3) are configured. So, no signalling is needed between LMA and MAG. Only, binding cache entries are updated at LMA and MAG and flows areforwarded according to policy rules on these entities. However, The LMA needs to know when to assign the same set of prefixes to all different physical interfaces of the mobile node. This can be achieved by different means, such as policy configuration. A new handover initiate (HI) value is defined to allow the MAG to indicate to the LMA that the same set of prefixes must be assigned to the mobile node.\\

\begin{figure}
	\centerline{\includegraphics[scale=0.5]{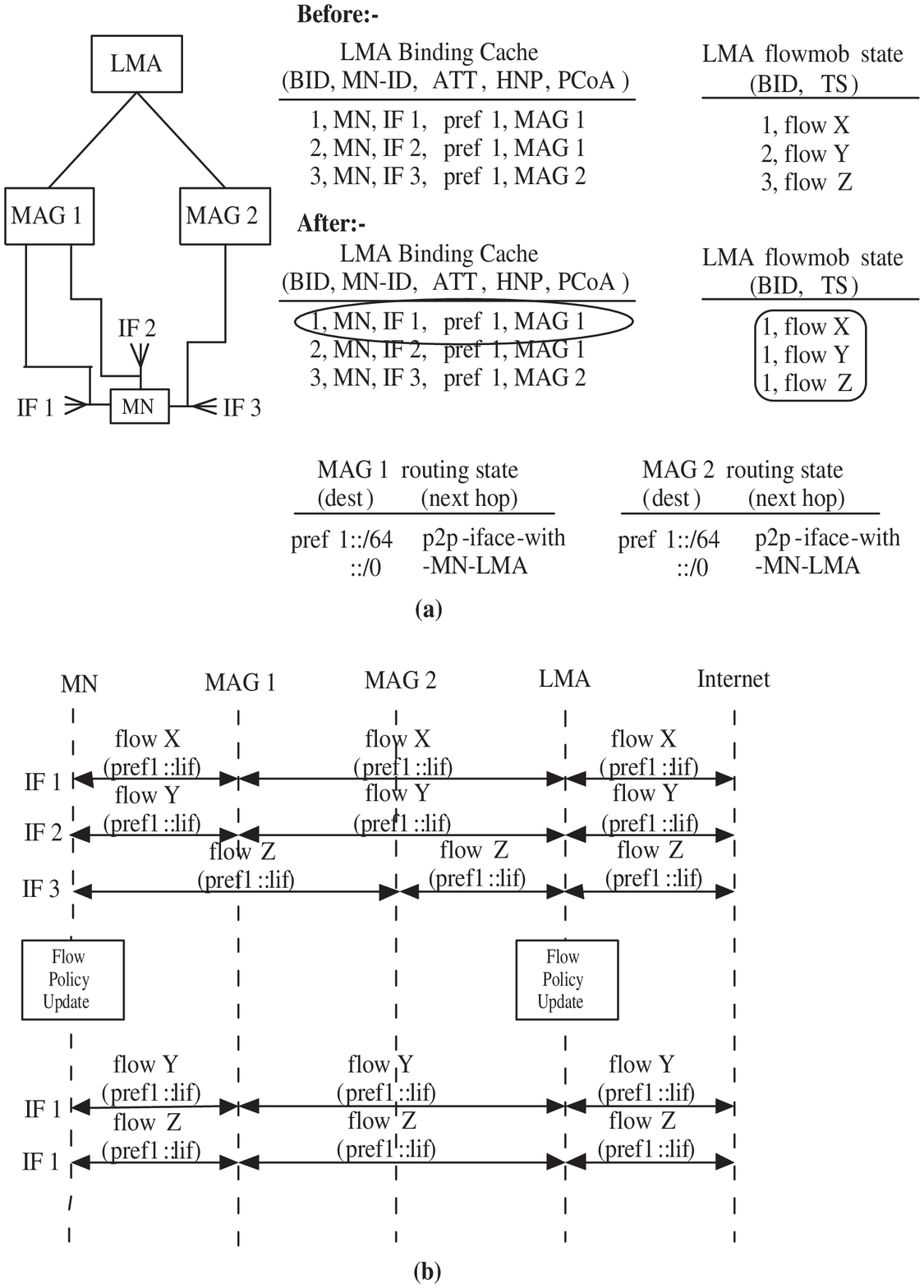}}
	\caption{When all interfaces are active and flows are using common prefix. (a)scenario with binding cache entries (b) signaling flow diagram}
	\label{ActiveShared}
\end{figure}

This is explained by considering a scenario shown in Fig. \ref{ActiveShared} with a signalling flow diagram and various binding cache entries. Initially, all the flows X, Y and Z are active on interface 1 (IF 1) and at a certain moment flow Y moved from interface 2 (IF 2) to IF 1 and uses the same prefix (pref 1) whereas flow Z moved from interface 3 (IF 3) to IF 1 and uses the same prefix (pref 1). From Fig. \ref{ActiveShared} (a), the LMA binding cache entries and flow mobility status are shown before and after the flow mobility. After moving the flows Y/Z from IF 2/IF 3 to IF 1, the corresponding flow mobility state is updated with the same binding ID (BID) 1 for X, Y and Z. From the signalling flow diagram, Fig. \ref{ActiveShared} (b), it is found that there is no signalling required to update the flow mobility status.
\subsubsection{Case 2: When all interfaces are active and flows use different prefix}
When all interfaces are active and flows use different prefixes. Initially, all the flows X, Y and Z are active on IF 1 and at a certain moment flow Y moved from IF 2 to IF 1 and uses a prefix 2 (pref 2), flow Z moved from IF 3 to IF 1 and uses a prefix 3 (pref 3). Here, a special signalling is required when a flow is to be moved from its original interface to a new one. Because, the LMA cannot send a proxy binding acknowledge (PBA) message that has not been triggered in response to a received proxy binding update (PBU) message. The trigger for the flow movement can be on mobile node (e.g., by using layer-2 signalling, by explicitly start sending flow packets via a new interface, etc.) or on the network (e.g., based on congestion and measurements performed at the network).\\

If the flow is being moved from its default path (which is determined by the destination prefix) to a different one, the LMA constructs a flow mobility initiate (FMI) message. This message must be sent to the new target MAG, i.e. the one selected to be used in the forwarding of the flow. LMA may send another FMI message, this time to remove the flow Z state at MAG 2. Otherwise the flow state at the MAG 2 will be removed upon timer expiration. However, the Flow Y can easily move to IF 1 and can use pref 2, it does not need any special signalling as MAG 1 routing state contains the pref 2 information. LMA will update binding cache about flow Y. So, after the flows Y/Z move from IF 2/IF 3 to IF 1, the corresponding flow mobility state is updated with same BID 1 for X, Y and Z. And, the LMA binding cache maintains all the prefixes of flows with same BID. Further, special signalling is now required to update the flow mobility status.
\begin{figure}
	\centerline{\includegraphics[scale=0.5]{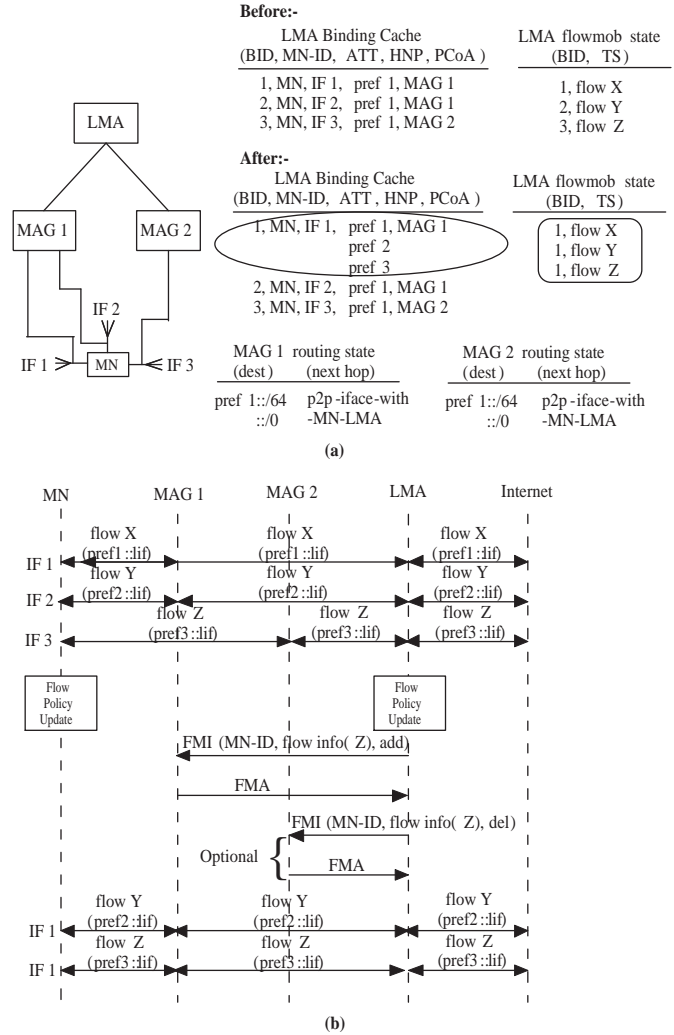}}
	\caption{When all interfaces are active and flows are using different prefix. (a) scenario with binding cache entries (b) signaling flow diagram}
	\label{Different}
\end{figure}
\subsubsection{Case 3: When some of the interfaces are suddenly powered on and flows use shared prefix}
When some of the interfaces are suddenly powered on and flows use shared prefix. Let, flows X, Y and Z be active on IF 1. At a certain moment, flow Y and flow Z detect congestion on IF 1 and MN powers on IF 2 and IF 3 and performs layer 2 (L2) attachment to MAG 1 and MAG 2 respectively. Upon L2 attachment of IF 2 to MAG 1 and IF 3 to MAG 2, a proxy binding registration is necessary. So, a proxy binding update message (PBU) will be sent to the LMA on behalf of IF 2 to the LMA, and a proxy binding acknowledgement (PBA) will be acknowledged back to IF 2 with pref 2. Similarly, for IF 3 the PBU will be sent from MAG 2 to LMA and the PBA will
be acknowledged along with pref 3 from the LMA. Then the corresponding binding cache entries will be updated at both the MAGs and LMA. Finally, the flow Y is moved to IF 2 and the flow Z is moved to IF 3 respectively. Here, the same pref 1 is shared among IF 2 and IF 3 for flow Y and flow Z. So, LMA moves pref 2 and pref 3 to new binding cache entry for IF 2 and IF 3 respectively. The LMA binding cache entries and flow mobility status are shown before and after the flow mobility. So, after the flows Y/Z move from IF 1 to IF 2/IF 3 respectively, the corresponding flow mobility state is updated with different BIDs 1, 2 and 3 for X, Y and Z respectively. And, the LMA binding cache maintains all the binding entries with same prefix (pref 1) to flows X, Y and Z. Now, proxy binding registrations are required for both IF 2 and IF 3 when these interfaces are suddenly powered on.
\begin{figure}
	\centerline{\includegraphics[scale=0.5]{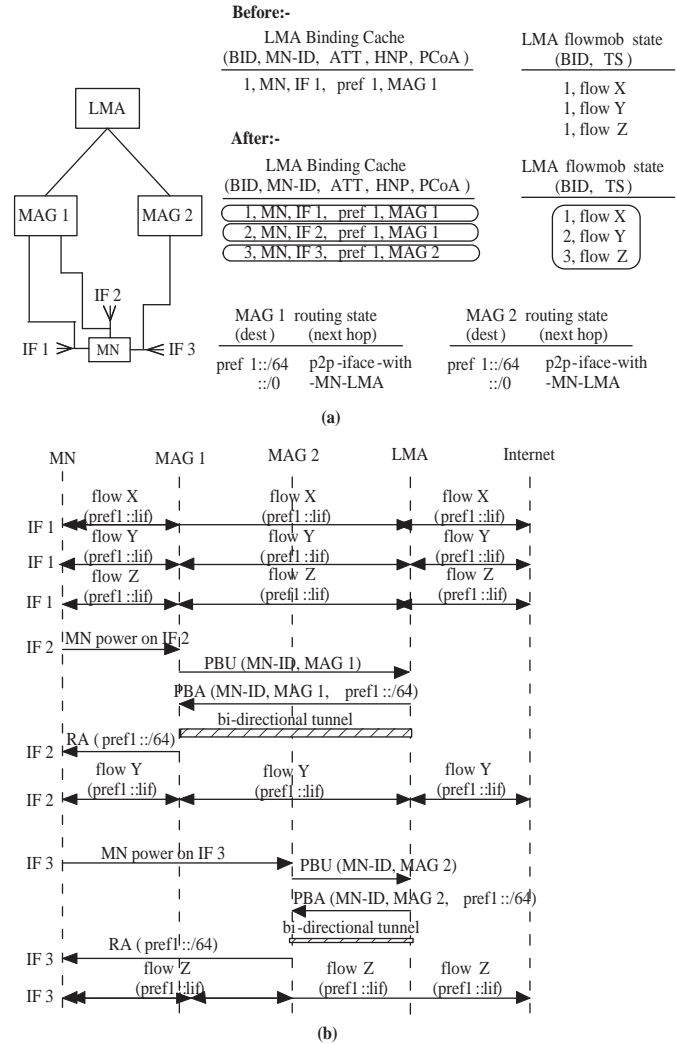}}
	\caption{When IF 2 and IF 3 suddenly power on and flows are using common prefix. (a)scenario with binding cache entries (b) signaling flow diagram}
	\label{NotactShared}
\end{figure}
\subsubsection{Case 4: When some of the interfaces are suddenly powered on and flows use different prefix}
Let flows X,Y and Z be active on IF 1. At a certain moment the flows Y and Z detect congestion on IF 1. So, the MN powers on IF 2 and IF 3 and performs L2 attachment to MAG 1 and MAG 2 respectively The flow Y is moved to IF 2 and the flow Z is moved to IF 3. When the MN powers on IF 2 and attaches to MAG 1, the LMA assigns pref 2 to flow Y and when the MN powers on IF 3 and attaches to the MAG 2, the LMA assigns pref 3 to flow Z. Finally, the LMA moves pref 2 to new binding cache entry for IF 2 and the LMA moves pref 3 to new binding cache entry for IF 3. The LMA binding cache entries and flow mobility status are shown before and after the flow mobility. So, after the flows Y/Z move from IF 1 to IF 2/IF 3 respectively, the corresponding flow mobility state is updated with different BIDs 1, 2 and 3 for X, Y and Z respectively. And, the LMA binding cache maintains all the binding entries with different prefixes (i.e. pref 1, pref 2, pref 3) to the flows X, Y and Z. Now proxy binding registrations are required for both IF 2 and IF 3 when these interfaces are suddenly powered on.\\

\begin{figure}
	\centerline{\includegraphics[scale=0.5]{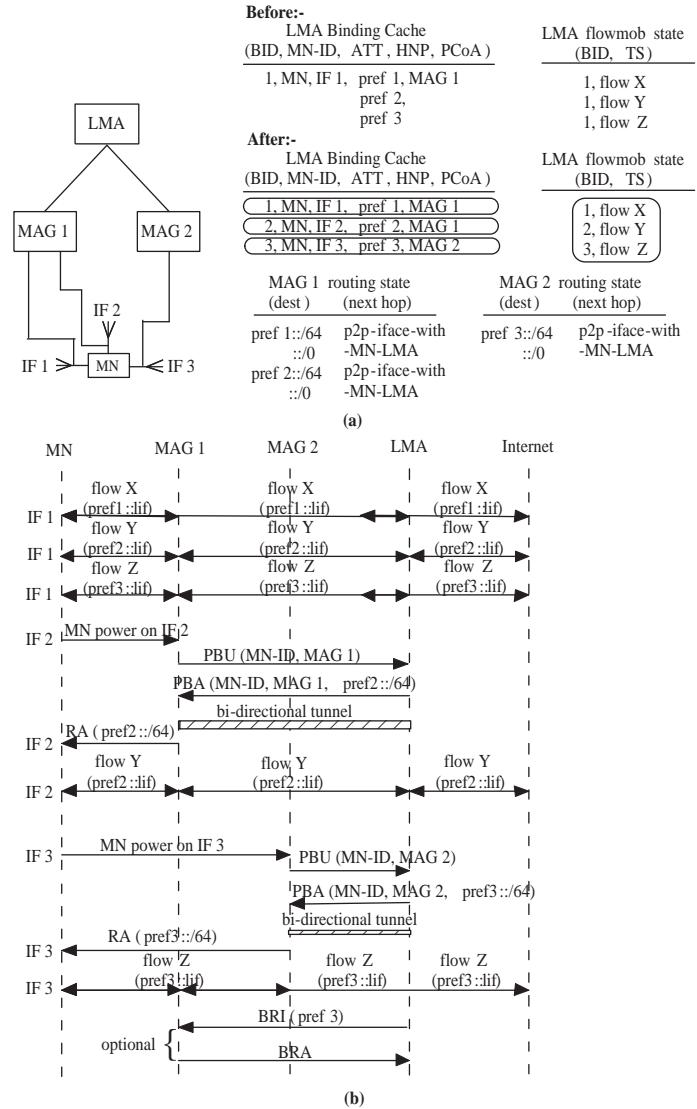}}
	\caption{When IF 2 and IF 3 suddenly power on and flows are using different prefix. (a)scenario with binding cache entries (b) signaling flow diagram}
	\label{NotactDifferent}
\end{figure}
\subsection{Multi LMA}
Flow mobility in multi-LMA can be seen for two different scenarios 1) when all the interfaces are active and when flow mobility takes place among the interfaces 2) when some of the interfaces are suddenly powered on and the flows need to be moved from
the currently active interfaces to newly active interfaces. Moreover, there are two different cases [7]: (i) when LMAs use a shared MAG, (ii) when LMAs use different MAGs. Thus, there are a total of four possible cases: (1) When all the interfaces active and LMAs use a shared MAG, (2) When all the interfaces active and LMAs use a different MAG, (3) When some of the interfaces suddenly powered on and LMAs use a shared MAG, (4) When some of the interfaces suddenly powered on and LMAs use a different MAG.
\subsubsection{Case1: When all the interfaces active and LMAs use a shared MAG}
\begin{figure}
	\centerline{\includegraphics[scale=0.5]{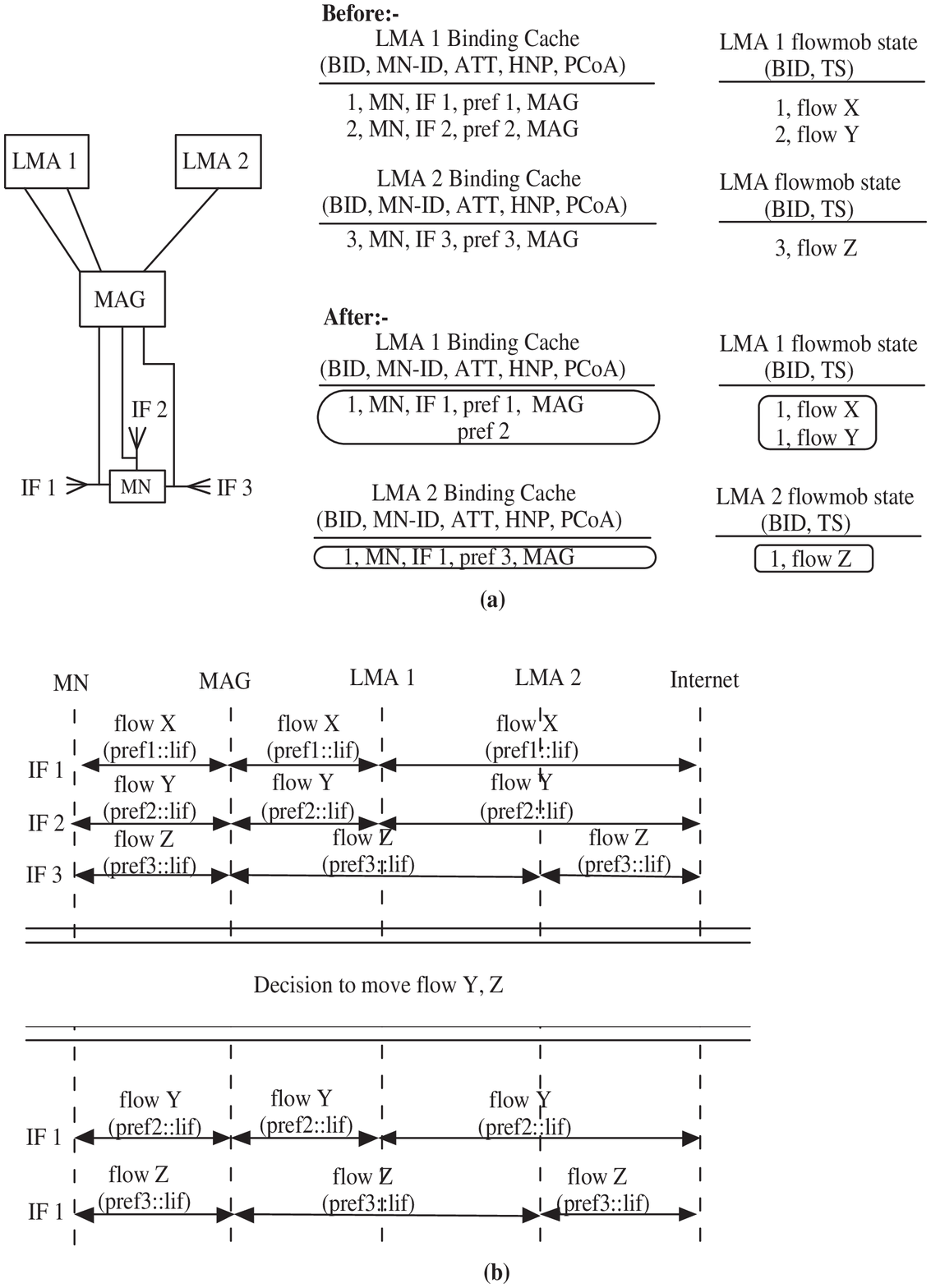}}
	\caption{When all the interfaces active and LMAs use a shared MAG. (a)scenario with binding cache entries (b) signaling flow diagram}
	\label{ActiveSharedMAG}
\end{figure}

When all the interfaces active and LMAs use a shared MAG. In this case no further signaling is required between LMA and MAG because routing entries are already present in the shared MAG. All physical interfaces are assigned with unique prefixes (i.e. pref 1, pref 2, and pref 3) upon attachment to the MAG. This is explained by considering a scenario shown in Figure \ref{ActiveSharedMAG}: When all interfaces are active. (a) scenario with binding cache entries (b) signaling flow diagram.\\

Initially, all the flows X, Y and Z are active on interface 1, 2, and 3 (IF 1, IF 2, and IF3) respectively and flows X, and Y attached to LMA1, and flow Z is attached to LMA2 through shared MAG. At a certain moment flows Y, Z detect congestion in the network and flow Y moved from interface 2 (IF 2) to IF 1 and uses the same prefix (pref 2) whereas flow Z moved from interface 3 (IF 3) to IF 1 and uses the same prefix (pref 3). As shown in Fig. \ref{ActiveSharedMAG} (a), before flowmobility operation, the binding cache entries of flows X/Y are at LMA1 and that of flow Z is at LMA2. As well as all the flows have different BIDs at their respective LMAs. So, after moving the flows Y/Z from IF 2/IF 3 to IF 1, the corresponding flow mobility state is updated with same binding ID (BID) 1 for X, Y and Z. However, the binding cache entries of the flows are updated in their respective LMAs only. From the signaling flow diagram Fig. \ref{ActiveSharedMAG} (b), it is shown that there is no signaling  required to update the flow mobility status.
\subsubsection{Case2: When all the interfaces active and LMAs use a different MAG}
Initially, all the flows X, Y and Z are active on IF 1, IF 2, and IF 3 and flows X, and Y attached to LMA1, and flow Z is attached to LMA2 through MAG1, and MAG2 respectively. At a certain moment flow Y moved from IF 2 to IF 1, flow Z moved from IF 3 to IF 1. Here, a special signaling is required when a flow is to be moved from its original interface to a new one. Because, the LMA 2 cannot send a proxy binding acknowledge (PBA) message that has not been triggered in response to a received proxy binding update
(PBU) message. If the flow Z is being moved from its default path (which is determined by the destination prefix) to a different one, the LMA 2 constructs a flow mobility initiate (eFMI) message. This message must be sent to the new target MAG, i.e. the
one selected to be used in the forwarding of the flow. This forwarding is done through LMA 1 as LMA 2 does not have any information about MAG 1 which is associated to the LMA 1. LMA 2 may send another FMI message, this time to remove the flow Z state at MAG 2. Otherwise the flow state at the MAG 2 will be removed upon timer expiration. However, the Flow Y can easily move to IF 1, it does not need any special signaling as MAG 1 routing state contains the pref 2 information. LMA 1 will update binding cache about flow Y. The LMA binding cache entries and flow mobility status is shown before and after flow mobility. So, after the flows Y/Z move from IF 2/IF 3 to IF 1, the corresponding flow mobility state is updated with same BID 1 for X, Y and Z.
\begin{figure}
	\centerline{\includegraphics[scale=0.5]{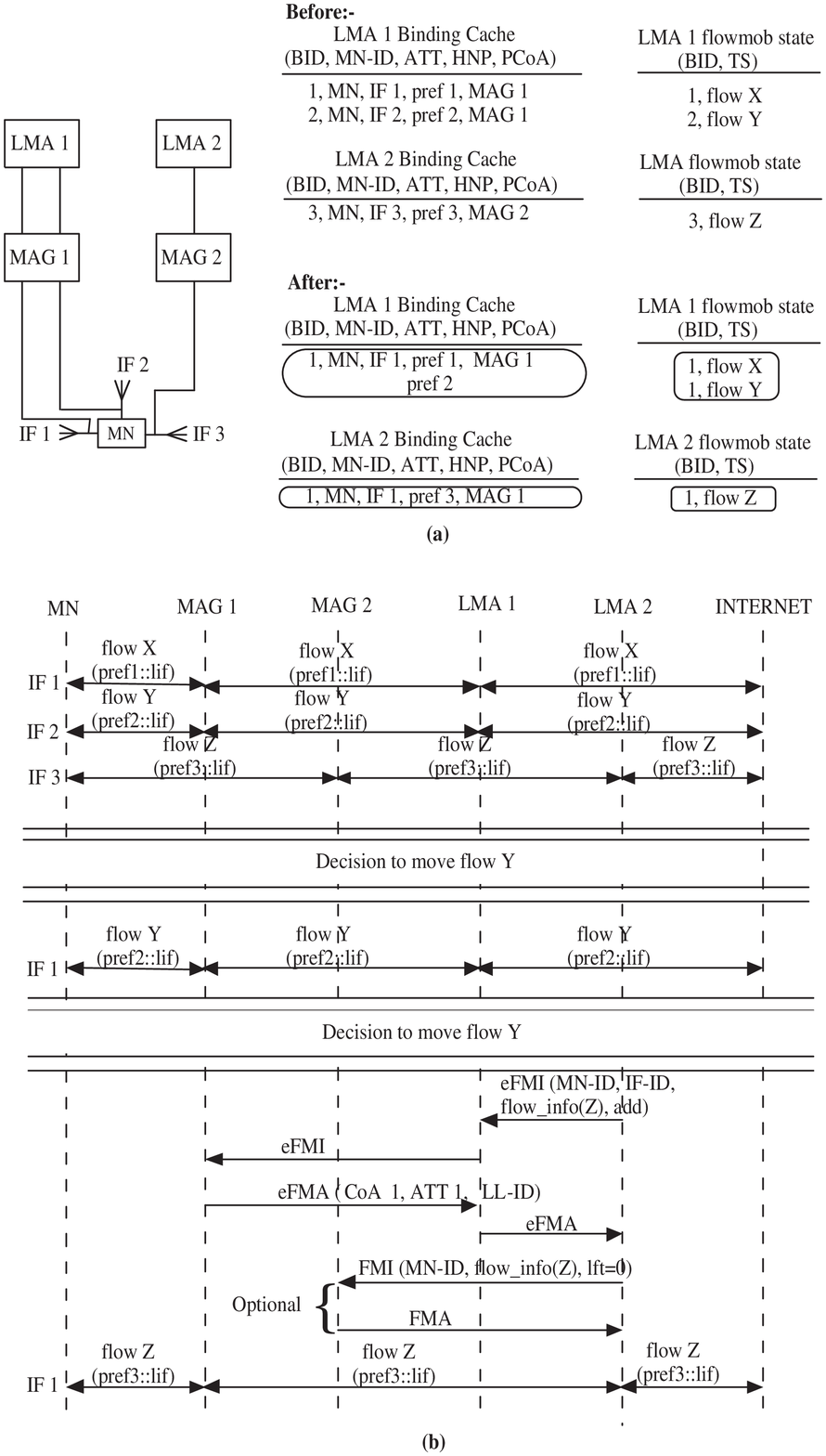}}
	\caption{When all the interfaces active and LMAs use a different MAG. (a)scenario with binding cache entries (b) signaling flow diagram}
	\label{ActiveDifferentMAG}
\end{figure}

\subsubsection{Case3: When some of the interfaces suddenly powered on and LMAs use a shared MAG}
where flows X, Y and Z are active on IF 1 and flows X, and Y attached to LMA1, and flow Z is attached to LMA2 through shared MAG. At a certain moment flow Y and flow Z detects congestion on IF 1 and MN powers on IF 2 and IF 3 perform layer 2 (L2) attachments to MAG. Upon L2 attachment of IF 2, and IF 3 to MAG, a proxy binding registration is necessary. So, a proxy binding update message (PBU) will be sent on behalf of IF 2 to the LMA 1, and a proxy binding acknowledgement (PBA) will be acknowledged back to IF 2 with pref 2. Similarly, for IF 3 the PBU will be sent from MAG to LMA 2 and the PBA will be acknowledged along with pref 3 from
the LMA 2. Then the corresponding binding cache entries will be updated at both the MAG and LMAs. Finally, the flow Y is moved to IF 2 and the flow Z is moved to IF 3 respectively. The LMA binding cache entries and flow mobility status
are shown before and after the flow mobility. So, after the flows Y/Z move from IF 1 to IF 2/IF 3 respectively, the corresponding flow mobility state is updated with different BIDs 1, 2 and 3 for X, Y and Z respectively. And, the LMA binding cache maintains all the binding entries with unique prefixes to flows X, Y and Z. Proxy binding registrations are required for both IF 2 and IF 3 when these interfaces are suddenly powered on.
\begin{figure}
	\centerline{\includegraphics[scale=0.5]{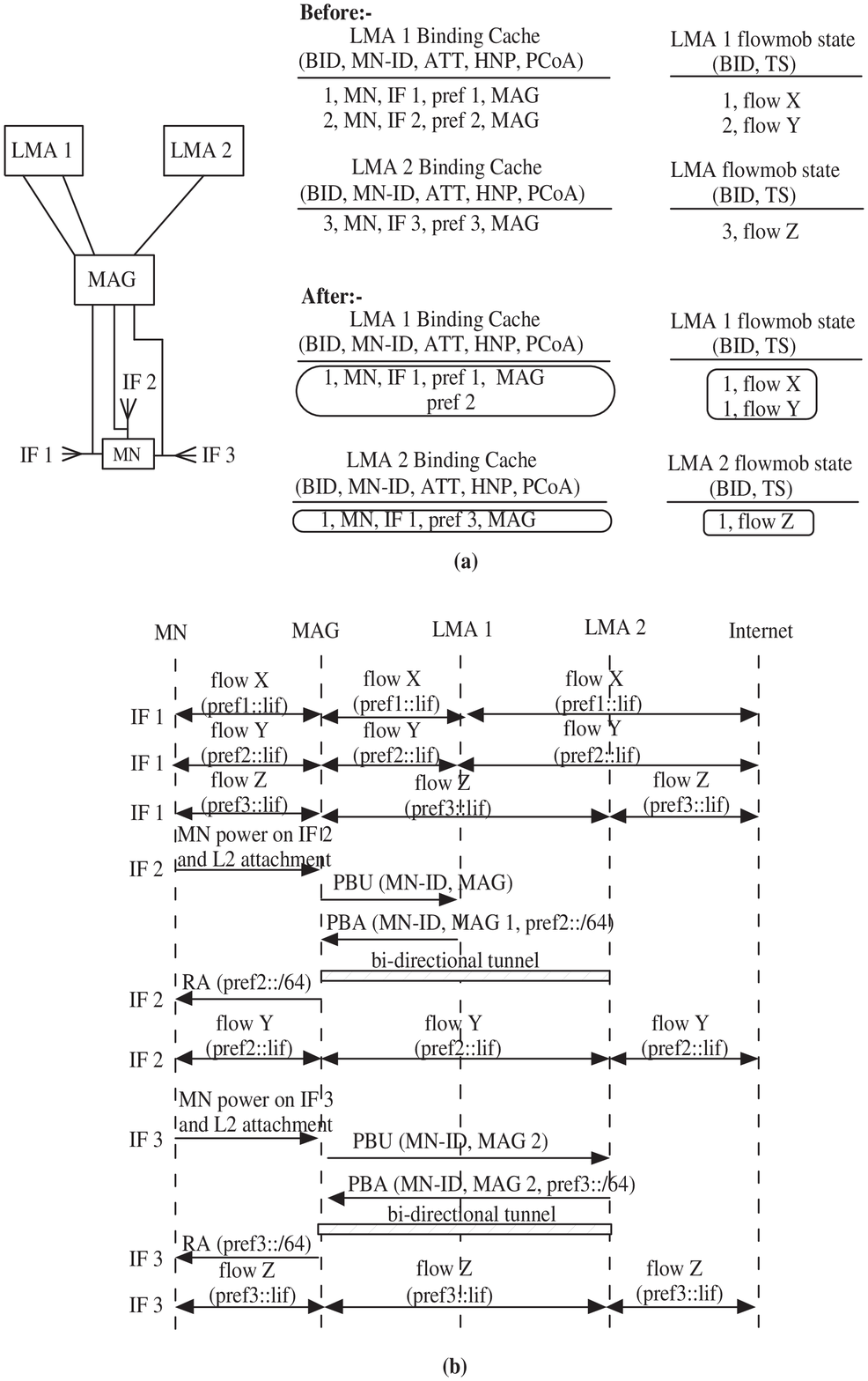}}
	\caption{When IF 2 and IF 3 suddenly powered on and LMAs use a shared MAG. (a)scenario with binding cache entries (b) signaling flow diagram}
	\label{NotActiveSharedMAG}
\end{figure}
\subsubsection{Case4: When some of the interfaces suddenly powered on and LMAs use a different MAG}
The signaling flow diagrams and various binding cache entries are explained as shown in Fig. 8, where flows X, Y and Z are active on IF 1 and flows X, and Y attached to LMA1, and flow Z is attached to LMA2 through MAG1, and MAG2 respectively. At a certain moment the flows Y and Z detects congestion on IF 1. So, MN powers on IF 2 and IF 3 and performs L2 attachment to MAG 1 and MAG 2 respectively. Then the flow Y is moved to IF 2 and the flow Z is moved to IF 3 respectively. When the MN powers on IF 2 and attaches to MAG 1, the LMA 1 assigns pref 2 to flow Y and when the MN powers on IF 3 and attaches to the MAG 2, the LMA 2 assigns pref 3 to flow Z. Finally, the LMA 1 moves pref 2 to new binding cache entry for IF 2 and the LMA 2 moves pref 3 to new binding cache entry for IF 3. The LMA binding cache entries and flow mobility status are shown before and after the flow mobility. So, after the flows Y/Z move from IF 1 to IF 2/IF 3 respectively, the corresponding flow mobility state is updated with different BIDs 1, 2 and 3 for X, Y and Z respectively. And, the LMA binding cache maintains all the binding entries with different prefixes (i.e. pref 1, pref 2, and pref 3) to the flows X, Y and Z. Proxy binding registrations are required for both IF 2 and IF 3 when these interfaces are suddenly powered on.\\

\begin{figure}
	\centerline{\includegraphics[scale=0.5]{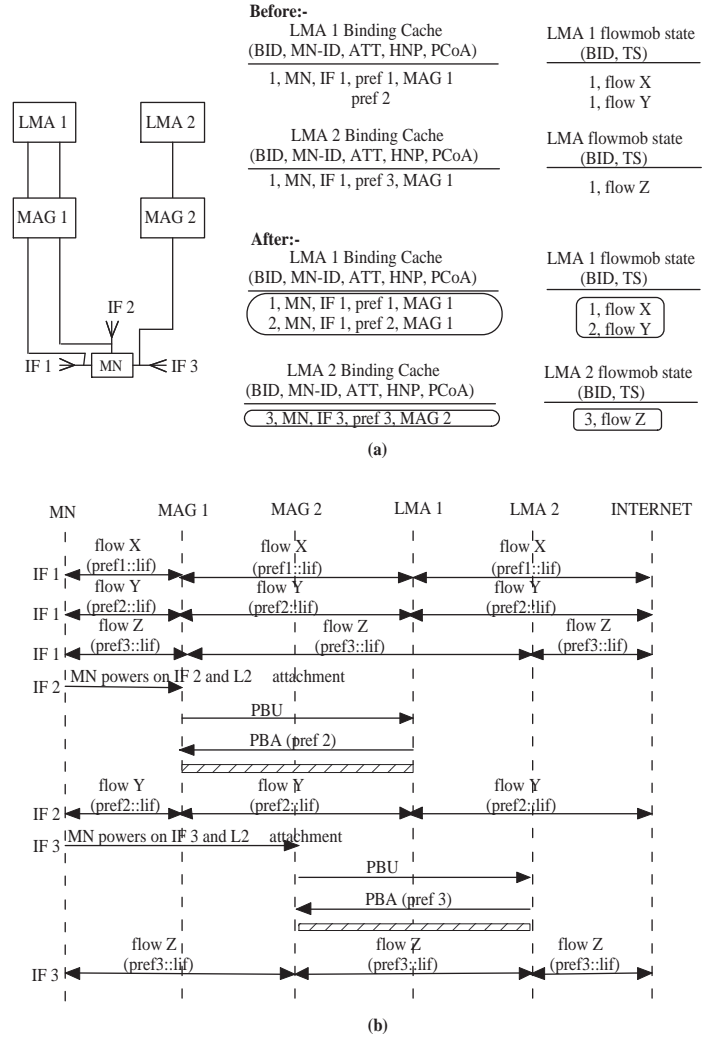}}
	\caption{When IF 2 and IF 3 suddenly powered on and LMAs use a different MAG. (a)scenario with binding cache entries (b) signaling flow diagram}
	\label{NotActiveDifferentMAG}
\end{figure}
From the above example scenarios in the presence of single LMA and multi LMA environments, for case 1 and case 2 , it has been observed that when all the interfaces are active, the flow can be provided with seamless mobility without any signalling.
However, it needs a special signalling when flows use different prefixes or LMAs use a different MAG in in single LMA, multi LMA environments respectively. But, keeping all the interfaces active when there is no network usage or when there are no flows
running, is not an effective solution. Because it leads to more battery drain up of the MN and more cost for the network usage. So, making the interfaces active dynamically or opportunistically based on the usage is more effective. From the above example scenarios, for case 3 and case 4, it has been observed that when some of the interfaces are suddenly powered on, and the flows are moving to newly active interfaces, it needs a special signalling to provide the seamless flow mobility. However, this extra signalling may cause performance degradation for the seamless flow mobility in terms of han-dover disruption delay, signalling cost or overhead, packet loss. The case becomes even worse when the interfaces of the MN increase. The reason behind extra signaling is lack of interface mobility, where the interface mobility defines that the flows of the same MN can use the different interfaces seamlessly. In the following section we will discuss our proposed block prefix mechanism to overcome these issues.	
	
\section{Proposed Block-Prefix Mechanism}
The proposed block prefix mechanism consists of two important phases of operation one is generation of home network block prefix (HNBP) from the prefixes assigned to different flows, and checking the home network prefixes (HNPs) of the flows at the MAG by using the stored HNBP. The flows active on current interface, using either same or unique prefix mechanisms, can use these prefixes even after moving to the newly attached interfaces of the MN. The HNBP generation is explained in Algorithm 1, where assuming that pref is the array containing the list of prefixes of active flows. The calculation of this HNBP from the prefixes (i.e. HNPs) of different active flows should be done either dynamically or periodically at LMA, and the same HNBP should be updated at both the LMA and the MAG.\\

After checking the HNPs of flows at MAGs while attaching through new interfaces, the binding or registration update status of flows should be done by sending an Update Status (US) message to the LMA. A new flag (B) is included in the proxy binding update message; rest of the proxy binding update message format remains the same as defined in \cite{RFC5213}. This new flag (B) is included in the proxy binding update message to indicate to the LMA that the flow is verified using a block prefix mechanism by the MAG. Value of B is 1 for an US message and must be set to 0 for a PBU message. The US message contains the interface ID along with the mobile node ID (MN-ID), which will be updated at the LMA. Due to this, it is possible to capture the interface mobility along with the flow mobility, which ultimately reduces the signalling involved during the flow mobility. As the flows belong to the same MN, but attach through different interfaces, only the binding entries need to be updated at the LMA; it does not require any acknowledgment and new prefixes to allow the flows. The flows can use the prefixes which were used before the flow mobility. However, the prefixes should be verified by the MAGs using the stored HNBP.\\

\begin{algorithm*}
	\caption{pseudocode for home network block prefix (HNBP) mechanism generation}
	\begin{algorithmic}
		\State $old_{-}prefix=pref;$ \Comment{Assign old prefix to current prefix}
		\State $HNBP=pref;$ \Comment{HNBP is the current prefix}
		\While{$pref \rightarrow next != null$}
		\State $new_{-}pref=pref \rightarrow next;$
		\If {$new_{-}pref==old_{-}pre$}
		\State $HNBP=old_{-}pref;$  \Comment{HNBP is the same prefix}
		\Else
		\State $HNBP=new_{-}pref + HNBP;$ \Comment{HNBP is the sum (logical OR) of all the prefixes of flows}
		\EndIf
		\EndWhile
	\end{algorithmic}
\end{algorithm*}
When all the interfaces are active, the flow mobility can take place among the interfaces. The flows can use either the same prefixes or unique prefixes after changing from one interface to another interface. For case 1, as discussed in Section 2, there is no need for any special signalling. Moreover, as in the case of a unique prefix mechanism (case 2 of Section 2), the special signalling such as FMI and FMA are also not required to intimate about the flow to the new target MAG. Because, the newly targeted MAG verifies the prefixes of the flows using the stored HNBP. However, there is an HNBP generation at LMA and HNP checking process at the MAGs, which need some extra processing cost; but, this does not contribute to any signalling overhead or cost. The HNBP generation process is done dynamically or periodically. The processing cost incurred due to this HNBP generation is negligible. Moreover, a special signalling such as US message is needed only when some of the interfaces suddenly become active i.e. as in the case 3 and case 4 of Section 2.
\subsection{LMA Operation}
Along with the basic operations of the LMA, it also has to support the following features:
1) LMA should generate the home network block prefix (HNBP) from the prefixes of the MN.
2) LMA should intimate the corresponding HNBP of MN to all the MAGs.
3) Upon the arrival of an update status (US) message, the LMA should be able to update the binding cache entries and interface-flow (if-flow) mobility cache of the flows and interfaces of the MN.
\subsection{MAG Operation}
Along with the basic operations of the MAG, it should be able to support the following features:
1) MAG should be able to store the received HNBP of MN from LMA.
2) Upon attachment of new interfaces, the MAG should check the prefixes of the newly attached flows, where these prefixes are verified by the stored HNBP of the MN.
3) After verification of flow's prefixes through new interfaces of the same MN, the MAG should send the update status message. This update status message consists of mobile node ID (MN-ID), and interface ID (IF-ID).
\subsection{Single LMA}
Assuming the LMA and MAG supports the block prefix mechanism with the above mentioned features, the proposed method can be explained with two different cases i.e. 1) when the interfaces are suddenly powered on and use the same prefixes 2) when the interfaces are suddenly powered on and use different prefixes.
\subsubsection{Case 1}
\begin{figure}
	\centerline{\includegraphics[scale=0.5]{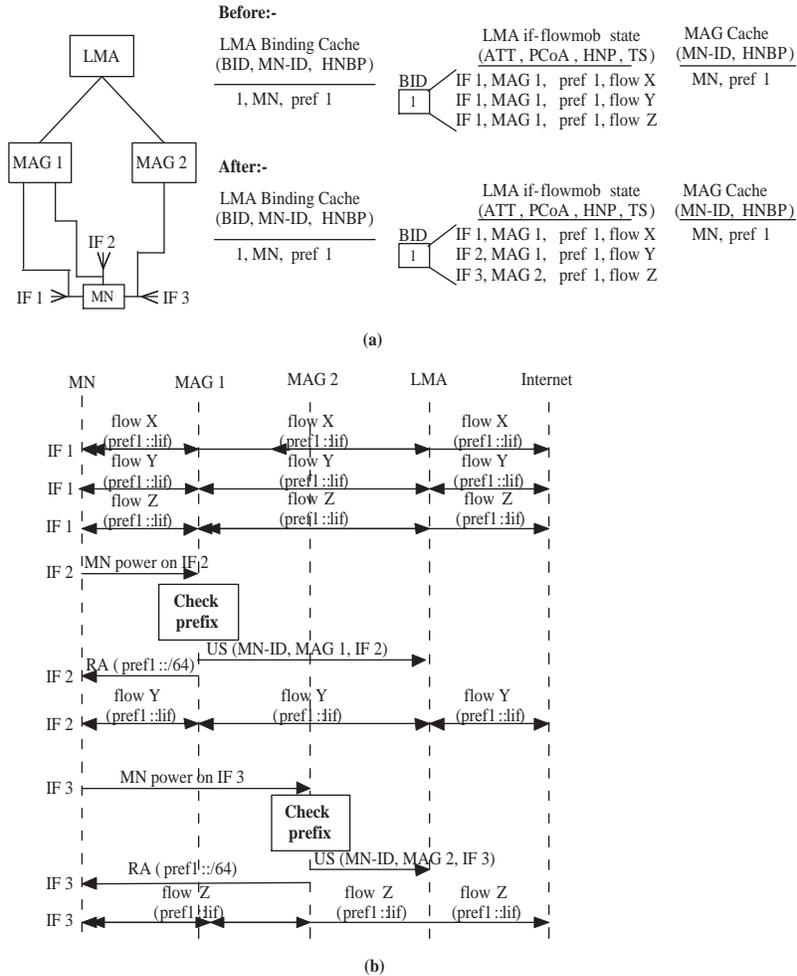}}
	\caption{When IF 2 and IF 3 suddenly power on, flows with common prefix and using block prefix mechanism. (a)scenario with binding cache entries (b) signaling flow}
	\label{NotactSharedBlock}
\end{figure}

Initially, all the flows X, Y and Z are active on IF 1 and sharing a common prefix (pref 1). Flow Y and flow Z detect congestion on IF 1 and MN powers on IF 2 and IF 3 and performs L2 attachment to MAG 1 and MAG 2 respectively. Based on the stored block prefix (HNBP), MAG 1 and MAG 2 have to decide the prefix authorization of flow Y and flow Z as given in Algorithm 2. Now, flow Y and flow Z can use the same prefix (pref 1). The signalling flow diagrams and various binding cache entries are shown in Fig. \ref{NotactSharedBlock}. In Fig. \ref{NotactSharedBlock} (a), the LMA binding cache entries and if-flow mobility status are shown before and after the flow mobility. So, after moving the flows Y and Z from IF 1 to IF 2 and IF 3 respectively, the corresponding if-flow mobility state is updated with the same BID 1 for all the flows X, Y and Z. Here, the respective interface IDs are also stored along with their flow IDs, which need to be updated in LMA if-flow mobility cache during the flow mobility. Moreover, in this particular case all flows are using the same prefix (pref 1) before and after the flow mobility. From the signalling flow diagram, Fig. \ref{NotactSharedBlock} (b), it is clear that when both IF 2 and IF 3 are suddenly powered on, update status messages are required for updating the if-flow mobility cache.
\begin{algorithm*}
	\caption{pseudocode for verification of flows when using common prefix}
	\begin{algorithmic}
		\If {$new_{-}pref \odot HNBP$}  \Comment{Check the new prefix at MAG using the stored HNBP}
		\State $Send_{-}US();$ \Comment{Allow flow Y and flow Z with same prefix}
		\Else
		\State $Proxy_{-}Binding_{-}Update();$  \Comment{Don't allow; need binding registration}
		\EndIf
	\end{algorithmic}
\end{algorithm*}
\subsubsection{Case 2}
\begin{figure}
	\centerline{\includegraphics[scale=0.5]{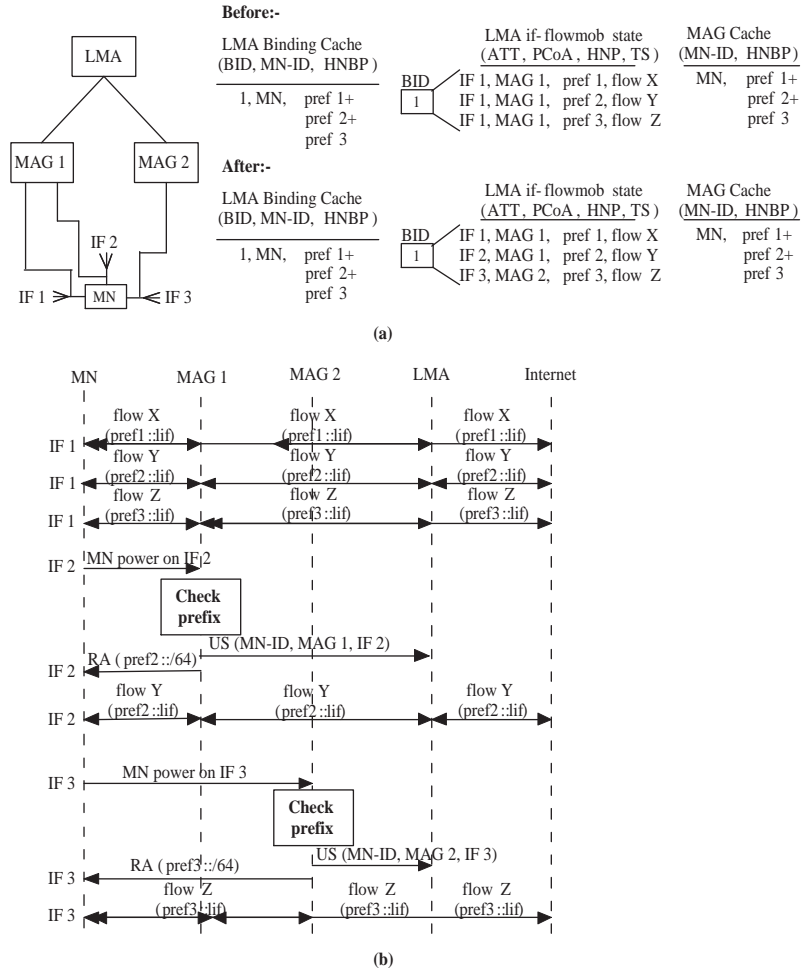}}
	\caption{When IF 2 and IF 3 suddenly power on, flows with different prefix and using block prefix mechanism. (a)scenario with binding cache entries (b) signaling flow}
	\label{NotactDifferentBlock}
\end{figure}

Initially, flows X, Y and Z are active on IF 1 using the unique prefixes pref 1, pref 2 and pref 3 respectively. Flows Y and Z detect congestion on IF 1 and MN powers on IF 2 and IF 3 and performs L2 attachment to MAG 1 and MAG 2 respectively. Based on the stored block prefix, MAG 1 and MAG 2 have to decide the prefix authorization of flow Y and flow Z as explained by Algorithm 3. Now, flow Y and flow Z can use the respective prefixes (i.e. pref 2 and pref 3). The signalling flow diagrams and various binding cache entries are explained in Fig. \ref{NotactDifferentBlock}. From Fig. \ref{NotactDifferentBlock} (a), the LMA binding cache entries and if-flow mobility status are shown before and after the flow mobility. So, after the flows Y and Z move from IF 1 to IF 2 and IF 3 respectively, the corresponding if-flow mobility state is updated with same BID 1 for all the flows X, Y and Z. Here, the respective interface IDs are also stored along with their flow IDs, which need to be updated in LMA if-flow mobility cache during the flow mobility. Moreover, in this particular case, all the flows are using unique prefixes (i.e. pref 1, pref 2 and pref 3) before and after the flow mobility. From the signalling flow diagram Fig. \ref{NotactDifferentBlock} (b), it is clear that when both IF 2 and IF 3 are suddenly powered on, update status messages are required for updating the if-flow mobility cache.
\begin{algorithm*}
	\caption{pseudocode for verification of flows (eg. flow Y and flow Z) when using different prefix}
	\begin{algorithmic}
		\State $Suppose~~HNBP~~is~~pref~1 + pref~2 + pref~3;$
		\If {$new_{-}pref \odot HNBP$}  \Comment{prefix verification at MAG 1 or MAG 2}
		\State $Send_{-}US();$ \Comment{Allow flow Y or Z with same prefix (pref 2 or pref 3) and send US to LMA}
		\Else
		\State $Proxy_{-}Binding_{-}Update();$  \Comment{Don't allow; need binding registration}
		\EndIf
	\end{algorithmic}
\end{algorithm*}

\subsection{Multi LMA}
Assuming the LMA and MAG supports the block prefix mechanism with the above mentioned features, the proposed method can be explained with two different cases i.e. 1) when the interfaces are suddenly powered on and use LMAs use shared MAG 2) when the interfaces are suddenly powered on and LMAs use different MAGs.
\subsubsection{Case 1: When some of the interfaces suddenly powered on and LMAs use a shared MAG}
\begin{figure}
	\centerline{\includegraphics[scale=0.5]{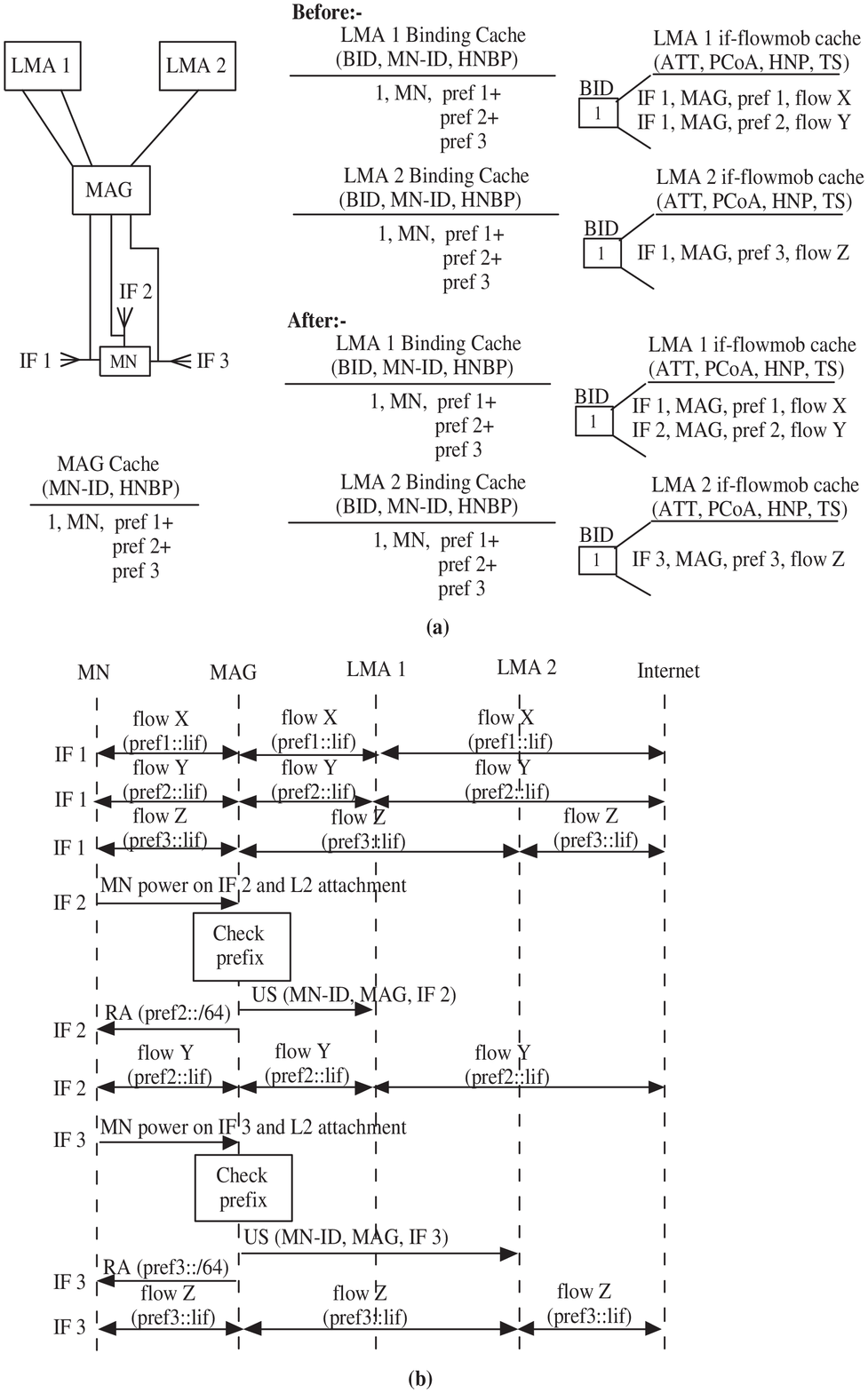}}
	\caption{When IF 2 and IF 3 suddenly powered on and LMAs use a shared MAG with block prefix. (a)scenario with binding cache entries (b) signaling flow}
	\label{NotActiveSharedMAGBlock}
\end{figure}

The signaling flow diagrams and various binding cache entries are explained as shown in Fig. \ref{NotActiveSharedMAGBlock}, where flows X, Y and Z are active on IF 1 and flows X, and Y attached to LMA1, and flow Z is attached to LMA2 through shared MAG. At a certain moment, flow Y and flow Z detects congestion on IF 1 and MN powers on IF 2 and IF 3 and perform layer 2 (L2) attachments to MAG. Based on the stored block prefix (HNBP), MAG has to decide the prefix authorization of flow Y and flow Z as given in Algorithm 2. The signaling flow diagrams and various binding cache entries are shown in Fig. \ref{NotActiveSharedMAGBlock}. In Fig. \ref{NotActiveSharedMAGBlock} (a), the LMA binding cache entries and if-flow mobility status are shown before and after the flow mobility. So, after moving the flows Y and Z from IF 1 to IF 2 and IF 3 respectively, the corresponding if-flow mobility state is updated with the same BID 1 for all the flows X, Y and Z. Here, the respective interface IDs are also stored along with their flow IDs, which need to be updated in LMA if-flow mobility cache during the flow mobility. From the signaling flow diagram, Fig. \ref{NotActiveSharedMAGBlock} (b), it is clear that when both IF 2 and IF 3 are suddenly powered on; update status messages are required for updating the if-flow mobility cache.

\subsubsection{Case 2: When some of the interfaces suddenly powered on and LMAs use a different MAG}
\begin{figure}
	\centerline{\includegraphics[scale=0.5]{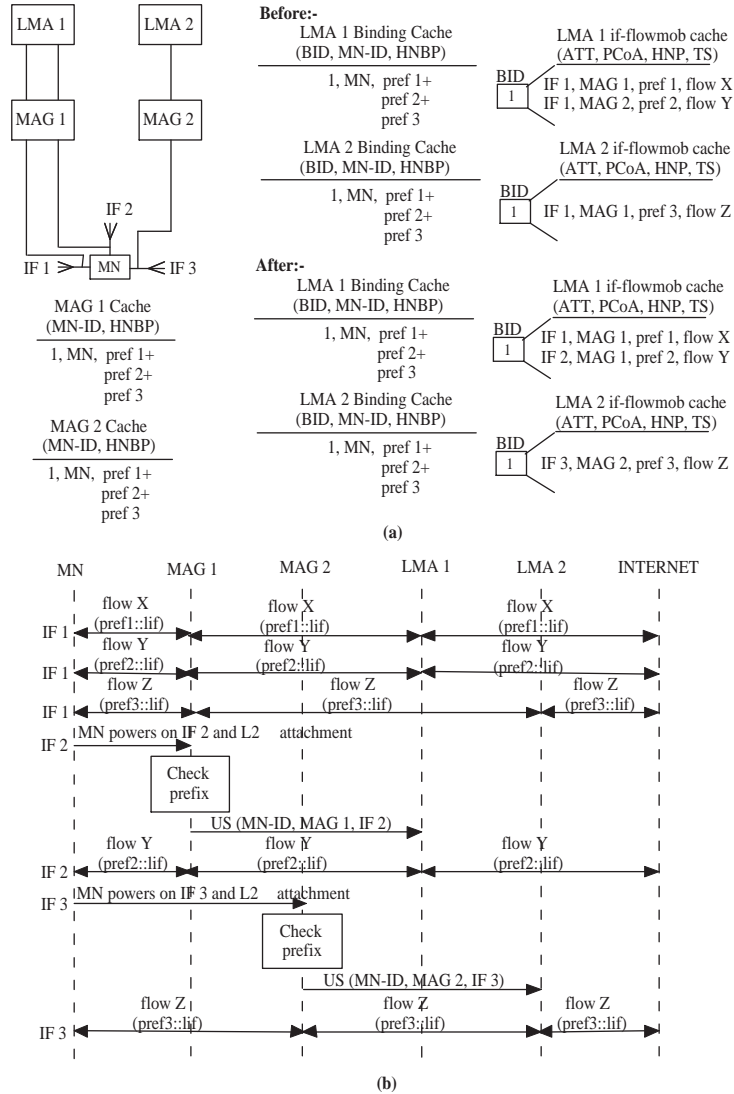}}
	\caption{When IF 2 and IF 3 suddenly powered on and LMAs use a different MAG with block prefix. (a)scenario with binding cache entries (b) signaling flow}
	\label{NotActiveDifferentMAGBlock}
\end{figure}

The signaling flow diagrams and various binding cache entries are explained as shown in Fig. \ref{NotActiveDifferentMAGBlock}, where flows X, Y and Z are active on IF 1 and flows X, and Y attached to LMA1, and flow Z is attached to LMA2 through MAG1, and MAG2 respectively. At a certain moment the flows Y and Z detects congestion on IF 1. So, MN powers on IF 2 and IF 3 and performs L2 attachment to MAG 1 and MAG 2 respectively. Then the flow Y is moved to IF 2 and the flow Z is moved to IF 3 respectively.  Based on the stored block prefix, MAG 1 and MAG 2 have to decide the prefix authorization of flow Y and flow Z as explained by Algorithm 3. The signaling flow diagrams and various binding cache entries are explained in Fig. \ref{NotActiveDifferentMAGBlock}. From Fig. \ref{NotActiveDifferentMAGBlock} (a), the LMA binding cache entries and if-flow mobility status are shown before and after the flow mobility. So, after the flows Y and Z move from IF 1 to IF 2 and IF 3 respectively, the corresponding if-flow mobility state is updated with same BID 1 for all the flows X, Y and Z. Here, the respective interface IDs are also stored along with their flow IDs, which need to be updated in LMA if-flow mobility cache during the flow mobility. From the signaling flow diagram Fig. \ref{NotActiveDifferentMAGBlock} (b), it is clear that when both IF 2 and IF 3 are suddenly powered on; update status messages are required for updating the if-flow mobility cache.\\

In the following section we will discuss the evaluation methodology, both analytically and through simulation of the proposed technique
\section{Analytical Evaluation and Simulation Study}
The network model of \cite{KiICC} is considered for handover latency analysis of flow mobility mechanisms in PMIPv6. In this model, $T^{X1}_{X2}$ denotes the delay due to the operation X2 of flow mobility technique X1 and $N^{X1}_{N1-N2}$ represents the number of hops between nodes N1 and N2 for the flow mobility technique X1. The symbols used to represent different delay variables are explained in Table \ref{Notations2}. Then, the total handover delay $D^{X1}_{HO}$ and the total number of hops $N^{X1}_{HO}$ during handover are derived for each of the technique.\\

\textbf{Single LMA}\\
As discussed in case 2 of single LMA environment in Section 2, from Fig. \ref{Different} (b), it is observed that it needs two signalling messages between LMA and MAG 1 and two messages between LMA and MAG 2. So, the total handover delay $D^{active_{-}diff}_{HO}$ is estimated as $2 \times t_{am}+2 \times t_{am}$. Similar analysis can be applied for the other techniques. Suppose, the technique X1 $\in \{active_{-}diff, notactive_{-}com, notactive_{-}diff, notactive_{-}com_{-}block, notactive_{-}diff_{-}block\}$ where these are as discussed in Case 2, Case 3 and Case 4 of single LMA environment in Section 2, Case 1 and Case 2 of the proposed block prefix mechanisms for single LMA environment respectively. Then, the average hop delay for a particular technique X1 is considered as the ratio of the total handover delay and number of hops during the handover interruption time, this is denoted as $Avg_{-}Hop_{-}Delay^{X1}_{HO}$ and are as follows:

\begin{align*}
D^{active_{-}diff}_{HO} & =2 \times t_{am}+2 \times t_{am}\\
N^{active_{-}diff}_{HO} & =2 \times N_{MAG_{-}LMA}+ 2 \times N_{MAG_{-}LMA}
\end{align*}
\begin{equation}
\label{ActiveDifferent}
Avg_{-}Hop_{-}Delay^{active_{-}diff}_{HO}=\frac{D^{active_{-}diff}_{HO}}{N^{active_{-}diff}_{HO}}
\end{equation}
\begin{align*}
D^{notactive_{-}com}_{HO} & =2 \times (t_{mr}+t_{ra})+4 \times t_{am}\\
N^{notactive_{-}com}_{HO} & =2 \times N_{MN_{-}MAG}+4 \times N_{MAG_{-}LMA}
\end{align*}
\begin{equation}
\label{NotactiveShared}
Avg_{-}Hop_{-}Delay^{notactive_{-}com}_{HO}=\frac{D^{notactive_{-}com}_{HO}}{N^{notactive_{-}com}_{HO}}
\end{equation}
\begin{align*}
D^{notactive_{-}diff}_{HO} & =2 \times (t_{mr}+t_{ra})+4 \times t_{am}+2 \times t_{am}\\
N^{notactive_{-}diff}_{HO} & =2 \times N_{MN_{-}MAG}+4 \times N_{MAG_{-}LMA}+2 \times N_{MAG_{-}LMA}
\end{align*}
\begin{equation}
\label{NotactiveDifferent}
Avg_{-}Hop_{-}Delay^{notactive_{-}diff}_{HO}=\frac{D^{notactive_{-}diff}_{HO}}{N^{notactive_{-}diff}_{HO}}
\end{equation}
\begin{align*}
D^{notactive_{-}com_{-}block}_{HO} & =2 \times (t_{mr}+t_{ra})+2 \times t_{am}\\
N^{notactive_{-}com_{-}block}_{HO} & =2 \times N_{MN_{-}MAG}+2 \times N_{MAG_{-}LMA}
\end{align*}
\begin{equation}
\label{NotactiveSharedBlock}
Avg_{-}Hop_{-}Delay^{notactive_{-}com_{-}block}_{HO}=\frac{D^{notactive_{-}com_{-}block}_{HO}}{N^{notactive_{-}com_{-}block}_{HO}}
\end{equation}
\begin{align*}
D^{notactive_{-}diff_{-}block}_{HO} & =2 \times (t_{mr}+t_{ra})+2 \times t_{am}\\
N^{notactive_{-}diff_{-}block}_{HO} & =2 \times N_{MN_{-}MAG}+2 \times N_{MAG_{-}LMA}
\end{align*}
\begin{equation}
\label{NotactiveDifferentBlock}
Avg_{-}Hop_{-}Delay^{notactive_{-}diff_{-}block}_{HO}=\frac{D^{notactive_{-}diff_{-}block}_{HO}}{N^{notactive_{-}diff_{-}block}_{HO}}
\end{equation}

\begin{table}[ht]
	\fontsize{7}{7pt}\selectfont
	\renewcommand{\arraystretch}{1}
	\caption{Symbols used for delay variables} 
	\centering 
	\begin{tabular}{c c } 
		\hline\hline 
		Delay & Simplified Notation \\ [0.5ex] 
		\hline 
		$T_{AP-MAG}$ & $t_{ra}$ \\ 
		$T_{MN-AP}$ & $t_{mr}$ \\
		$T_{MAG-LMA}$ & $t_{am}$ \\ [1ex] 
		\hline 
	\end{tabular}
	\label{Notations2} 
\end{table}
\textbf{Multi LMA}\\
As discussed in case 2 of multi LMA environment in Section 2, from Fig. \ref{Different} (b), it is observed that it needs two signalling messages between MAG and LMA 1 and two messages between MAG and LMA 2. So, the total handover delay $D^{active_{-}2MAG}_{HO}$ is estimated as $2 \times t_{am}+2 \times t_{am}$. Similar analysis can be applied for the other techniques. Suppose, the technique X1 $\in \{active_{-}2MAG, notactive_{-}1MAG, notactive_{-}2MAG, notactive_{-}1MAG_{-}block, notactive_{-}\\2MAG_{-}block\}$ where these are as discussed in Case 2, Case 3 and Case 4 of multi LMA environment in Section 2, Case 1 and Case 2 of the proposed block prefix mechanisms for multi LMA environment respectively. Then, the average hop delay for a particular technique X1 is considered as the ratio of the total handover delay and number of hops during the handover interruption time, this is denoted as $Avg_{-}Hop_{-}Delay^{X1}_{HO}$ and are as follows:
\begin{align*}
D^{active_{-}2MAG}_{HO} & =2 \times t_{pn}+2 \times t_{am}+2 \times t_{am}\\
N^{active_{-}2MAG}_{HO} & =2 \times N_{LMA_{-}LMA}+ 2 \times N_{MAG_{-}LMA}+2 \times N_{MAG_{-}LMA}
\end{align*}
\begin{equation}
\label{ActiveDifferent}
Avg_{-}Hop_{-}Delay^{active_{-}2MAG}_{HO}=\frac{D^{active_{-}2MAG}_{HO}}{N^{active_{-}2MAG}_{HO}}
\end{equation}
\begin{align*}
D^{notactive_{-}1MAG}_{HO} & =2 \times (t_{mr}+t_{ra})+4 \times t_{am}\\
N^{notactive_{-}1MAG}_{HO} & =2 \times N_{MN_{-}MAG}+4 \times N_{MAG_{-}LMA}
\end{align*}
\begin{equation}
\label{NotactiveShared}
Avg_{-}Hop_{-}Delay^{notactive_{-}1MAG}_{HO}=\frac{D^{notactive_{-}1MAG}_{HO}}{N^{notactive_{-}1MAG}_{HO}}
\end{equation}
\begin{align*}
D^{notactive_{-}2MAG}_{HO} & =2 \times (t_{mr}+t_{ra})+4 \times t_{am}\\
N^{notactive_{-}2MAG}_{HO} & =2 \times N_{MN_{-}MAG}+4 \times N_{MAG_{-}LMA}+2 \times N_{MAG_{-}LMA}
\end{align*}
\begin{equation}
\label{NotactiveDifferent}
Avg_{-}Hop_{-}Delay^{notactive_{-}2MAG}_{HO}=\frac{D^{notactive_{-}2MAG}_{HO}}{N^{notactive_{-}2MAG}_{HO}}
\end{equation}
\begin{align*}
D^{notactive_{-}1MAG_{-}block}_{HO} & =2 \times (t_{mr}+t_{ra})+2 \times t_{am}\\
N^{notactive_{-}1MAG_{-}block}_{HO} & =2 \times N_{MN_{-}MAG}+2 \times N_{MAG_{-}LMA}
\end{align*}
{\small
	\begin{equation}
	\label{NotactiveSharedBlock}
	Avg_{-}Hop_{-}Delay^{notactive_{-}1MAG_{-}block}_{HO}=\frac{D^{notactive_{-}1MAG_{-}block}_{HO}}{N^{notactive_{-}1MAG_{-}block}_{HO}}
	\end{equation}}
\begin{align*}
D^{notactive_{-}2MAG_{-}block}_{HO} & =2 \times (t_{mr}+t_{ra})+2 \times t_{am}\\
N^{notactive_{-}2MAG_{-}block}_{HO} & =2 \times N_{MN_{-}MAG}+2 \times N_{MAG_{-}LMA}
\end{align*}
{\small
	\begin{equation}
	\label{NotactiveDifferentBlock}
	Avg_{-}Hop_{-}Delay^{notactive_{-}2MAG_{-}block}_{HO}=\frac{D^{notactive_{-}2MAG_{-}block}_{HO}}{N^{notactive_{-}2MAG_{-}block}_{HO}}
	\end{equation}}
\subsection{Handover Latency}
A better mobility management protocol is the one that consists of minimum number of hops with minimum signalling packet size. Assuming an M/M/1 queuing model, the packet service delay $D_{P}$ at each node is given by
$$D_{P}=\frac{1}{(1-\rho)*\mu}$$
where $\mu$ is the mean service rate at each node and $\rho=\frac{\lambda}{\mu}$. Then, the total handover delay during the flow mobility is given by (\ref{Delay})\cite{VasuPM2HW2N}
\begin{equation}
\label{Delay}
T_{d}=\frac{\lambda*H*(T_{h}+D_{P})}{V_{f}*(1-\frac{K}{K_{Max}})}
\end{equation}
where $\lambda$ is the packet arrival rate, $T_{h}$ is the average hop delay calculated above, and the $\frac{K}{K_{Max}}$ is the packet density ratio.
\subsection{Signalling cost}
The work in \cite{MingIEEETOC} clearly describes that the radio channel characteristics predominate the link performance for slower mobile nodes, while node mobility dominates the link performance for faster mobile nodes and also shown that the link dwell time can be effectively approximated by an exponential distribution. So, the link holding time of the mobile node is assumed to follow an exponential distribution with mean value of $\frac{1}{\mu_{L}}$ and sessions are assumed to arrive according to a Poisson process with a mean value of $\lambda_{S}$. It is assumed that there is no message transmission failure other than in the wireless link and also that the wired link is robust with respect to packet loss. $p_{f}$ is the probability of wireless link failure. If the probability density function is given as $f^{*}_{L}(t)$, then its Laplace transform is given by $$f^{*}_{L}(s)=\int^{\infty}_{t=0}e^{-st}f^{*}_{L}(t)dt$$
For $N_{L}$ number of link changes during inter-session time interval, the probability that the mobile node moves across K links is given by $Pr[N_{L}=K]=\alpha(K)$ (\cite{JongIEEETVT}, \cite{JongIEEETOCE}). So, We have $$\alpha(0)=1-\frac{1-(f^{*}_{L}(\lambda_{S}))}{S_{\sigma}}$$ and $$\alpha(K\geq1)=\frac{1}{S_{\sigma}} \times (1-(f^{*}_{L}(\lambda_{S}))^{2}) \times (f^{*}_{L}(\lambda_{S}))^{2}$$ where $S_{\sigma}$ is the session to mobility ratio (SMR) and is given by $\frac{\lambda_{S}}{\mu_{L}}$. The signalling cost is defined as the mobility signalling overhead incurred during a flow mobility management operation. It is assumed that the size of message M is $S_{M}$, $M \in \{RS, RA, PBU, PBA, FMI, FMA, BRI, BRA, US\}$. The signalling cost for protocol X  is denoted as $C_{X}$ and message overhead cost as $OH_{X}$, and are given for all the flow mobility management techniques, as follows:
\begin{align*}
C_{X} & = [i\Sigma^{\infty}_{i=0}\alpha(i)] \times OH_{X}\\
& = [\Sigma^{\infty}_{i=0}\frac{i}{S_{\sigma}} \times (1-f_{L}^{*}(\lambda_{S})) \times (f_{L}^{*}(\lambda_{S}))^{i-1}] \times OH_{X}
\end{align*}
\textbf{Single LMA}
\begin{align*}
OH_{active_{-}diff} & =N^{active_{-}diff}_{MAG-LMA} \times (S_{FMI}+S_{FMA})+N^{active_{-}diff}_{MAG-LMA} \times (S_{FMI}+S_{FMA})
\end{align*}

\begin{align*}
OH_{notactive_{-}com} & =(\frac{p_{f}}{1-p_{f}}) \times 2 \times S_{RA}+2 \times N^{notactive_{-}com}_{MAG-LMA} \times (S_{PBU}+S_{PBA})+\\
&2 \times (N^{notactive_{-}com}_{MAG-MN}-1) \times S_{RA}
\end{align*}

\begin{align*}
OH_{notactive_{-}diff} & =(\frac{p_{f}}{1-p_{f}}) \times 2 \times S_{RA}+2 \times N^{notactive_{-}diff}_{MAG-LMA} \times (S_{PBU}+S_{PBA})+\\
&2 \times (N^{notactive_{-}diff}_{MAG-MN}-1)\times S_{RA}+2 \times N^{notactive_{-}diff}_{MAG-LMA} \times (S_{BRI}+S_{BRA})
\end{align*}

\begin{align*}
OH_{notactive_{-}com_{-}block} & =(\frac{p_{f}}{1-p_{f}}) \times 2 \times S_{RA}+2 \times N^{notactive_{-}com_{-}block}_{MAG-LMA} \times (S_{US})+\\
&2 \times (N^{notactive_{-}com_{-}block}_{MAG-MN}-1) \times S_{RA}
\end{align*}

\begin{align*}
OH_{notactive_{-}diff_{-}block} & =(\frac{p_{f}}{1-p_{f}}) \times 2 \times S_{RA}+2 \times N^{notactive_{-}diff_{-}block}_{MAG-LMA} \times (S_{US})+\\
&2 \times (N^{notactive_{-}diff_{-}block}_{MAG-MN}-1) \times S_{RA}
\end{align*}
\textbf{Multi LMA}
\begin{align*}
OH_{active_{-}2MAG} & =N^{active_{-}2MAG}_{MAG-LMA} \times (S_{eFMI}+S_{eFMA})+N^{active_{-}2MAG}_{LMA-LMA} \times (S_{eFMI}+S_{eFMA})\\
&+N^{active_{-}2MAG}_{LMA-LMA} \times (S_{eFMI}+S_{eFMA})
\end{align*}

\begin{align*}
OH_{notactive_{-}1MAG} & =(\frac{p_{f}}{1-p_{f}}) \times 2 \times S_{RA}+2 \times N^{notactive_{-}1MAG}_{MAG-LMA}\times (S_{PBU}+S_{PBA})+ \times \\
&(N^{notactive_{-}1MAG}_{MAG-MN}-1) \times S_{RA}
\end{align*}

\begin{align*}
OH_{notactive_{-}2MAG} & =(\frac{p_{f}}{1-p_{f}}) \times 2 \times S_{RA}+2 \times N^{notactive_{-}2MAG}_{MAG-LMA} \times (S_{PBU}+S_{PBA})+\\
&2 \times (N^{notactive_{-}2MAG}_{MAG-MN}-1)\times S_{RA}
\end{align*}

\begin{align*}
OH_{notactive_{-}1MAG_{-}block} & =(\frac{p_{f}}{1-p_{f}}) \times 2 \times S_{RA}+2 \times N^{notactive_{-}1MAG_{-}block}_{MAG-LMA} \times (S_{US})+\\
&2 \times (N^{notactive_{-}1MAG_{-}block}_{MAG-MN}-1) \times S_{RA}
\end{align*}

\begin{align*}
OH_{notactive_{-}2MAG_{-}block} & =(\frac{p_{f}}{1-p_{f}}) \times 2 \times S_{RA}+2 \times N^{notactive_{-}2MAG_{-}block}_{MAG-LMA} \times (S_{US})+\\
&2 \times (N^{notactive_{-}2MAG_{-}block}_{MAG-MN}-1) \times S_{RA}
\end{align*}
\subsection{Packet Loss}
In a network, packet loss can occur due to various factors such as variations in signal strength, bad channel effects, congestion, corrupted packets, failed nodes, and queuing effects etc. Even though bit error rate (BER) is considered a QoS parameter in (\cite{3GPP_IMT_QoS_2002}, \cite{3GPP_TS_23.107}), in heterogeneous wireless networks it is not sufficient to decide the application quality. Rather, packet loss can be a useful parameter to decide the application level QoS. Many schemes have been proposed in the literature to reduce the packet loss during handover. However, it is important to note that packet loss does not always indicate a problem while it is tolerable. Suppose, the packets are arriving with an arrival rate of $\lambda$ during the flow mobility management procedure; the number of packets lost ($P_{L}$) is the product of packet arrival rate and the disruption time and is give by (\ref{packetloss})
\begin{equation}
\label{packetloss}
P_{L}=\lambda \times T_{d}
\end{equation}
\noindent where $T_d$ is the link disruption time.

\subsection{Simulation methodology}
Packets are generated using a Poisson process with a mean value of 100 pkts/sec and delay between various nodes such as $t_{mr}$, $t_{ra}$ and $t_{am}$ are assumed to be uniformly distributed from the mean values given in Table \ref{Assumptions1} \cite{KiICC} within the range of $mean \pm 50\%mean$. At each node packets are being serviced with a service rate of 150 pkts/sec. Using simulation, packet density is calculated as the number of packets arrived per unit hop during the handover. The simulation is run for 10000 distributed values of packet arrivals, and different delay variables such as $t_{mr}$, $t_{ra}$ and $t_{am}$. The total simulation is run for 1000 times and the average values are obtained. Various performance measures such as total handover delay, average hop delay, packet density, packet arrival rate are calculated from the simulations. The values are used to evaluate the performance relations among the parameters. The simulation results are matching qualitatively with the analytical result.
In the following section we will present both analytical and simulation results obtained using the proposed model.
\section{Results and Discussions}
This section compiles results of numerical analysis and simulation studies of various flow mobility management techniques in PMIPv6 under the assumptions presented in Table \ref{Assumptions1} \cite{KiICC} regarding the protocol operation delay and number of hops involved. The message overheads considered for different kind of signalling messages are shown in Table \ref{Assumptions2} (\cite{JongIEEETVT}; \cite{JongIEEETOCE}]. When all the interfaces are active and using techniques such as common prefix or block prefix mechanism for the flows, it does not need any signalling other than policy updates. So, the corresponding results are not given in the results and discussion section.\\

\begin{table}[ht]
	\vspace{-0.3cm}
	\fontsize{7}{9pt}\selectfont
	\renewcommand{\arraystretch}{1}
	\caption{Assumptions} 
	\centering 
	\begin{tabular}{c c } 
		\hline\hline 
		Number of Hops & Delay \\ [0.5ex] 
		\hline 
		$MN-MAG=MAG-MN=2$ & $t_{mr}=10ms$ \\ 
		$MAG-LMA=2$ & $t_{ra}=2ms$ \\
		$MAG-MAG=1$ & $t_{am}=20ms$ \\ [1ex] 
		\hline 
	\end{tabular}
	\label{Assumptions1} 
\end{table}
\begin{table}[ht]
	\fontsize{7}{7pt}\selectfont
	\renewcommand{\arraystretch}{1}
	\caption{Message Over Head} 
	\centering 
	\begin{tabular}{c c } 
		\hline\hline 
		Message Type & Size (Bytes) \\ [0.5ex] 
		\hline 
		$S_{RS}$ & $80$ \\ 
		$S_{RA}$ & $90$ \\
		$S_{PBU}$ & $76$ \\
		$S_{PBA}$ & $76$ \\
		$S_{FMI}$ & $56$ \\
		$S_{FMA}$ & $56$ \\
		$S_{US}$ & $56$\\
		$S_{BRI}$ & $56$ \\
		$S_{BRA}$ & $56$ \\ [1ex] 
		\hline 
	\end{tabular}
	\label{Assumptions2} 
\end{table}

\begin{figure}
	\centerline{\includegraphics[scale=0.5]{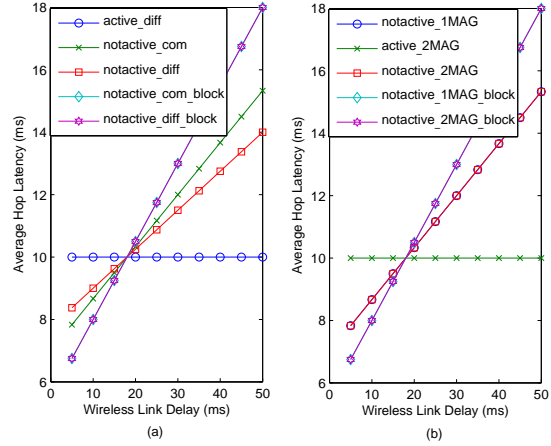}}
	\caption{Average Hop Latency(msec) vs. Wireless Link Delay(msec) for different mechanisms: Active Different, Nonactive Common, Nonactive Different, Block Prefix with Shared, Block Prefix with Different. (a) Single LMA (b) Multi LMA}
	\label{WirelessLinkDelay}
\end{figure}

As handover latency is directly proportional to average hop delay, it is important to study the effect of average hop delay. From Fig. \ref{WirelessLinkDelay} (a) and (b), it is observed that when all interfaces are active and each of the flows uses a unique prefix in single LMA case and multi LMA case with multiple MAGs, the average hop delay does not change with respect to wireless link delay. Because, in this case even though it needs a special signaling for flow mobility, it does not participate in the radio access signalling procedure. However, when some of the interfaces are suddenly active and use either common or different prefixes in single LMA and multi LMA case, the average hop delay increases with the increase in the wireless link delay due to the radio access involvement in signalling. Here, we observe a cross-over point (i.e. at the junction of all the techniques) below which a radio access technology can be treated as faster technology and above which a radio access can be treated as slower technology. For faster radio access technology, the average hop delay gives better performance for the block prefix mechanism compared to other techniques. However, the performance is opposite above the cross-over point (i.e. for slower radio access technologies), which means that the performance for block prefix mechanism is poorer compared to the other technique. Moreover, performance is the same when flows adopt the common or unique prefixes which using the block prefix mechanism. But, the performance is different and better when using unique prefix than common prefix in single LMA case where as it is same when using single MAG or multiple MAG in multi LMA case.\\

\begin{figure}
	\centerline{\includegraphics[scale=0.5]{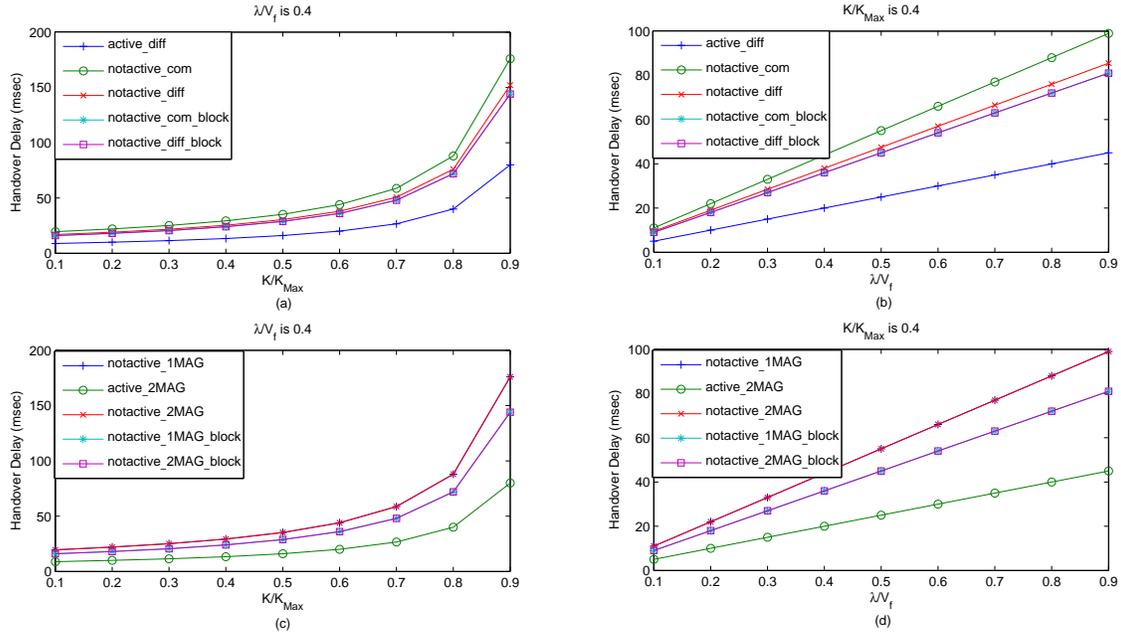}}
	\caption{Handover latency vs. packet density ($\frac{K}{K_{Max}}$) (a) Single LMA (c) Multi LMA; Handover latency vs. packet arrival rate ($\frac{\lambda}{V_{f}}$) (b) Single LMA (d) Multi LMA}
	\label{HandoverLatency}
\end{figure}

Fig. \ref{HandoverLatency}(a) and (c) shows the variation in handover latency versus the packet density (K). From this figure, it is observed that for a particular $\frac{\lambda}{V_{f}}$, the handover latency increases non-linearly with $\frac{K}{K_{Max}}$ for a high packet density. This is due to a large number of binding updates and signalling overhead. When applying the block prefix mechanism to the flows while using common or different prefixes in sling LMA case or using single MAG or multiple MAGs in multi LMA case, the handover delay is less compared to the other two mechanisms. This is due to a reduction in hop delay product involved. The variation of handover latency with the packet arrival rate ($\lambda$) is shown in Fig. \ref{HandoverLatency}(b) and (d). From this figure, it is observed that for a particular $\frac{K}{K_{Max}}$ value, as $\frac{\lambda}{V_{f}}$ increases, handover latency increases linearly; this is due to the increase in number of arrivals for binding updates and signalling overhead during the handover. When applying block prefix mechanism, the performance in terms of handover latency is better compared to the other two mechanisms. This is due to the reduction in the number of signalling operations. However, it is less compared to the case where all the interfaces are suddenly active and use unique prefixes for flows in sling LMA case or using either single MAG or multiple MAG in multi LMA case.\\

\begin{figure}
	\centerline{\includegraphics[scale=0.5]{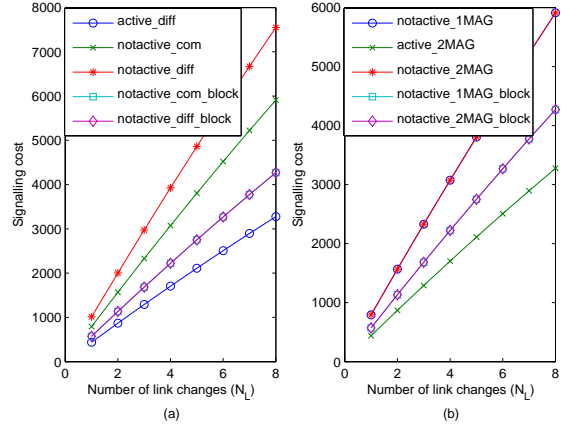}}
	\caption{Signalling cost vs. Number of link changes ($N_{L}$) for different mechanisms. (a) Single LMA (b) Multi LMA}
	\label{CostLinkChanges}
\end{figure}

As the number of link changes increases, the signaling cost increases linearly as shown in Fig. \ref{CostLinkChanges}(a) and (b). Because, the mobile node needs to send binding information to the new network, which causes the mobility management protocols to handle the related signalling. When some of the interfaces are suddenly active and use either common or different prefixes in single LMA case or using single MAG or multiple MAG in multi LMA case, the signalling cost will be more due to a higher signaling overhead incurred to update the LMA binding cache and the flow mobility state. But, the signalling cost is same in case of multi LMA case whereas it is more for unique prefix method in single LMA case. However, if we apply the block prefix mechanism with either common or different prefixes in single LMA case or using single MAG or multiple MAGs in multi LMA case, the performance will be better than that with the other two mechanisms. Moreover, the signalling cost does not depend on the type of prefixes we are using (i.e. shared or unique) as the block prefix mechanism maintains the interface mobility along with the flow mobility, this reduce the signaling overhead.\\

\begin{figure}
	\centerline{\includegraphics[scale=0.5]{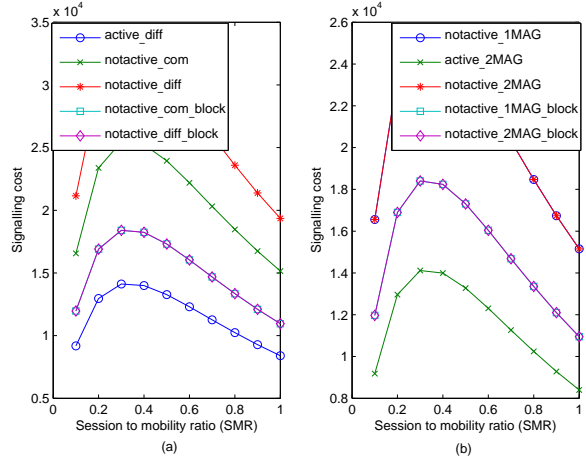}}
	\caption{Signalling cost vs. Session to mobility ratio (SMR) for different mechanisms. (a) Single LMA (b) Multi LMA}
	\label{CostSMR}
\end{figure}

As shown in Fig. \ref{CostSMR} (a) and (b), with an increase in the session to mobility ratio, the signalling cost of flow mobility management protocols decreases. Because the SMR is high for a low mobility of mobile nodes, implying that the chance of activation of flow mobility management operations is less, which ultimately reduces the total signalling overhead. However, there is a threshold point of SMR for a fixed number of link changes. From the graph, it is observed that the threshold is 0.3 for three link changes. As the number of link changes and SMR range varies, the threshold point may also vary.\\

\begin{figure}
	\centerline{\includegraphics[scale=0.5]{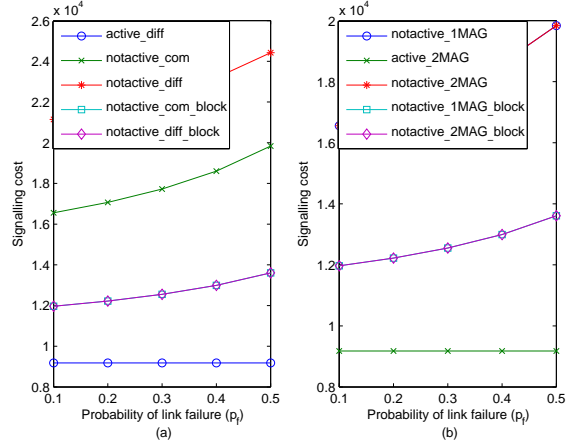}}
	\caption{Signalling cost vs. Probability of link failure ($p_{f}$) for different mechanisms. (a) Single LMA (b) Multi LMA}
	\label{CostLinkFailure}
\end{figure}

The variation in signalling cost versus the probability of link failure is shown in Fig. \ref{CostLinkFailure} (a) and (b). It is observed that the rate of change of signalling cost is more when flows are using either common or different prefixes in single LMA case or using single MAG or multiple MAGs in multi LMA case. However, the rate of change in signalling cost for block prefix mechanism, either with common or unique prefixes, is low due to its less radio access involvement. Because, most of the signalling is done through network entities on behalf of the mobile node. Moreover, the signalling cost is same when applying block prefix mechanism to the flows that use either common or unique prefixes.\\

\begin{figure}
	\centerline{\includegraphics[scale=0.5]{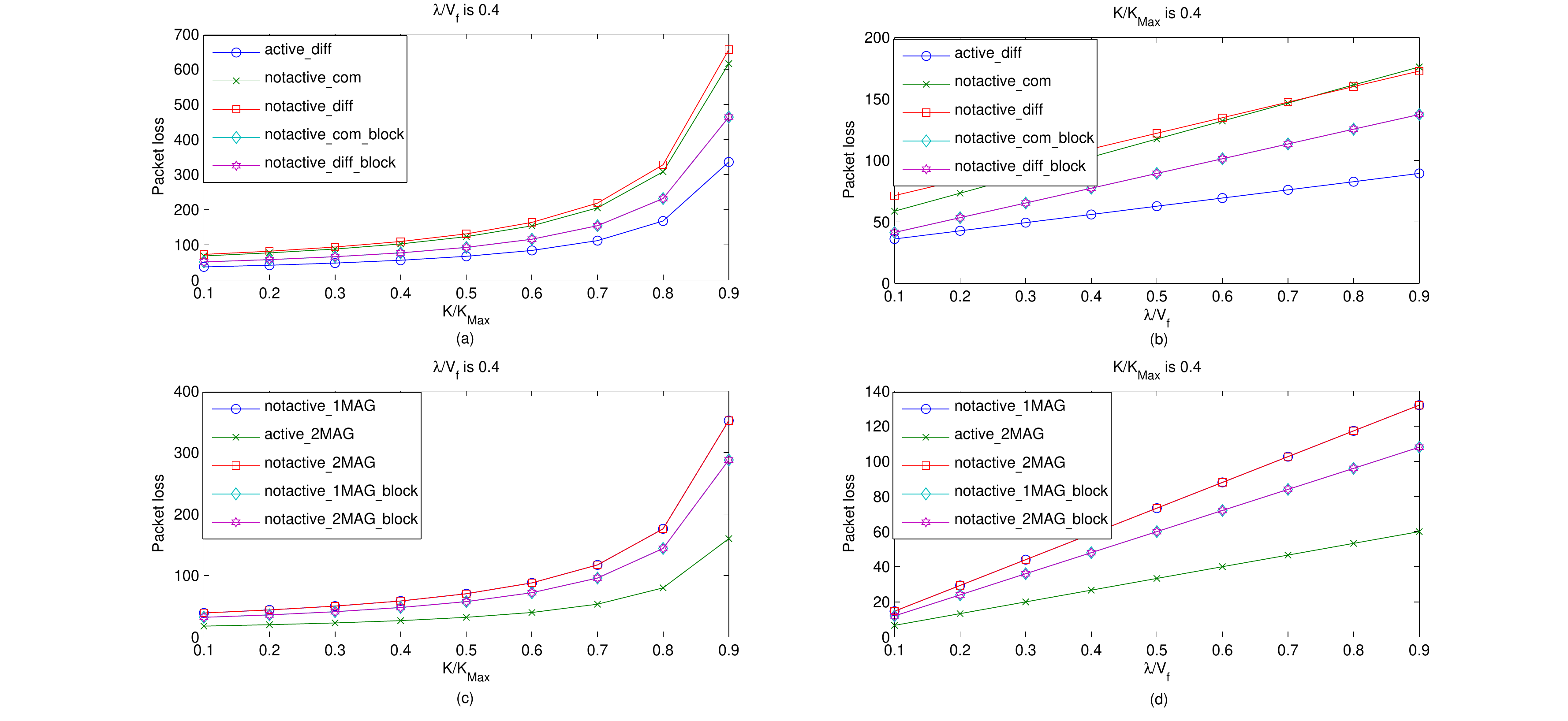}}
	\caption{Packet loss vs. packet arrival rate ($\frac{\lambda}{V_{f}}$) (a) Single LMA (c) Multi LMA; Packet loss vs.packet density ($\frac{K}{K_{Max}}$) (b) Single LMA (d) Multi LMA}
	\label{PacketLoss}
\end{figure}

The packet loss performance in terms of packet density and packet arrival rate is explained in Fig. \ref{PacketLoss}(a) and (c) and Fig. \ref{PacketLoss}(b) and (d), respectively. For a particular packet density ratio ($\frac{K}{K_{Max}}$), as $\frac{\lambda}{V_{f}}$ increases the number of packets lost increases linearly. And, for a particular $\frac{\lambda}{V_{f}}$ ratio, as packet density ratio ($\frac{K}{K_{Max}}$) increases the packet loss increases non-linearly. As packet density increases, the congestion or delay increase in the intermediate hops, increasing the packet loss rate. A similar interpretation regarding the techniques can be applied above in the case of handover latency analysis.\\

\begin{figure}
	\centerline{\includegraphics[scale=0.5]{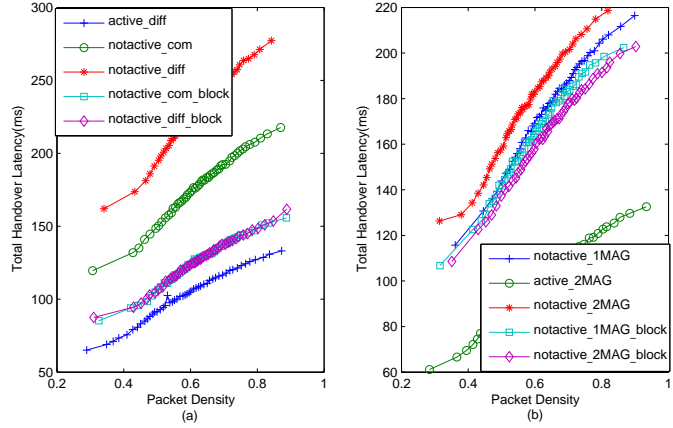}}
	\caption{Total Handover Latency (msec) vs. packet density ($\frac{K}{K_{Max}}$) for different mechanisms. (a) Single LMA (b) Multi LMA}
	\label{SimulationK}
\end{figure}

\begin{figure}
	\centerline{\includegraphics[scale=0.5]{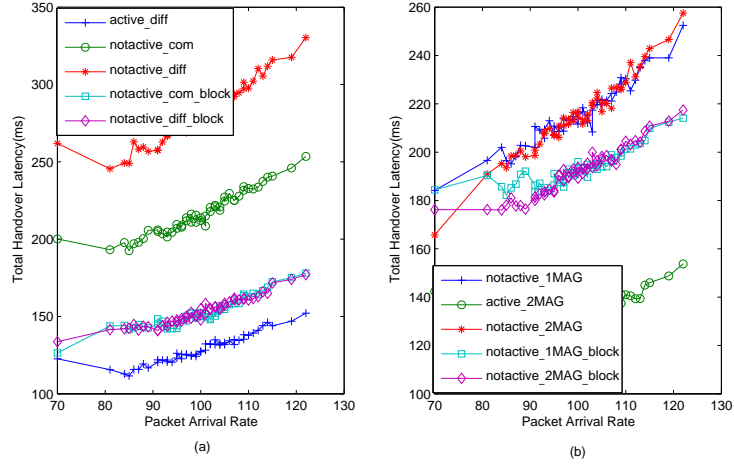}}
	\caption{Total Handover Latency (msec) vs. packet arrival rate ($\lambda$) for different mechanisms. (a) Single LMA (b) Multi LMA}
	\label{SimulationPkt}
\end{figure}

Simulations studies were also carried out to observe the variation in handover latency with the packet density ($\frac{K}{K_{Max}}$) and the results are depicted in Fig. \ref{SimulationK} (a) and (b). From the figure it is observed that as the packet density increases the total handover latency increases non-linearly. This matches with the analytical results given in Fig. \ref{HandoverLatency}(a) and (c). Simulation results depicted in Fig. \ref{SimulationPkt} (a) and (b), which shows the variation in handover latency with the packet arrival rate ($\lambda$). From this Fig. \ref{SimulationPkt} (a) and (b), it is observed that for a particular $\frac{K}{K_{Max}}$ value, the increase in handover latency is almost linear in nature. A similar observation is made from the analytical results, shown in Fig. \ref{HandoverLatency} (b) and (d). Moreover, it is observed that when applying the block prefix mechanism either using common or different prefixes in the flows, the handover disruption time in term of packet density, and packet arrival rate is less compared to without block prefix mechanism.\\


From the analytical and simulation results it is observed that the proposed block prefix mechanism performs better compared to the existing flow mobility management procedures that either use shared or unique prefixes in single LMA case or single MAG or multiple MAG in multi LMA case. The key reason for this is that while using the Block prefix mechanism, the interface mobility status is also captured along with the flow mobility status, which ultimately reduces the signaling involved for the flow mobility management.

\section{Conclusion}
In this paper, a new block prefix mechanism is developed and applied for flow mobility management in PMIPv6 networks. Analysis is done in terms of various performance metrics like handover latency, average hop delay, packet density, signalling cost, and packet loss. Both analytical and simulation results demonstrate that the proposed block prefix mechanism performs better compared to the existing flow mobility management procedures either using shared or unique prefixes. Because, while using block prefix mechanism, the interface mobility status is also captured along with the flow mobility status, which ultimately reduces the signaling involved for the flow mobility management. Moreover, our study also considered flow mobility management in Multi-LMA based networks in addition to those with a single LMA.



\end{document}